%% file: PorrmannBartelsVoigt_v3.tex
\documentclass[preprint, showkeys,onecolumn,nofootinbib,superscriptaddress]{revtex4-2}

\usepackage{amsthm,
  amsmath,
  amssymb,
  enumitem,
  geometry,
  caption,
  graphicx,
  fontenc,
  inputenc}
\usepackage[hidelinks]{hyperref}
\usepackage{amsbsy, orcidlink}
\newcommand{\R}{\mathbb{R}}
\renewcommand{\vec}[1]{\boldsymbol{#1}}
\newcommand{\normal}{\vec{\nu}}
\renewcommand{\S}{\mathcal{S}}
\renewcommand{\H}{\mathcal{H}}
\newcommand{\K}{\mathcal{K}}
\newcommand{\baseManifold}{\mathcal{M}}
\newcommand{\param}{\vec{X}}
\newcommand{\B}{\boldsymbol{B}}
\renewcommand{\P}{\boldsymbol{P}}
\newcommand{\Id}{\textbf{Id}}
\newcommand{\tangent}{\mathrm{T}}
\newcommand{\paramUpdate}{\vec{Y}}
\newcommand{\paramUpdateTest}{\vec{Z}}
\newcommand{\flow}{\vec{u}}
\newcommand{\flowTest}{\vec{v}}
\newcommand{\HTest}{\mathcal{G}}
\newcommand{\pressure}{p}
\newcommand{\pressureTest}{q}
\newcommand{\stress}{\vec{\sigma}}
\newcommand{\pos}{\vec{x}}
\newcommand{\posy}{\vec{y}}
\newcommand{\Reynolds}{\text{Re}}
\newcommand{\Bending}{\text{Be}}
\newcommand{\friction}{\gamma}
\newcommand{\areagrowth}{\delta_{\mathrm{A}}}

\newcommand{\Helfrich}{F_{\mathrm{B}}}
\newcommand{\TP}{\Phi}
\newcommand{\TPexponent}{\alpha}
\newcommand{\tr}{\operatorname{tr}}

\newcommand{\gradS}{\nabla_{\!\S}}
\newcommand{\divS}{\operatorname{div_{\!\S}}}
\newcommand{\gradC}{\operatorname{\nabla_{\!C}}}
\newcommand{\divC}{\operatorname{div_{\!C}}}
\newcommand{\DivC}{\operatorname{Div_{\!C}}}

\newcommand{\lapS}{\operatorname{\Delta_{\S}}}
\newcommand{\lapC}{\operatorname{\Delta_{C}}}
\newcommand{\deformation}[1]{\eth_{#1}}
\newcommand{\areaelement}[1]{d\S \ifthenelse{\equal{#1}{}}{}{_{#1}}}
\newcommand{\Lone}{\mathrm{L}^1}
\newcommand{\Ltwo}{\mathrm{L}^2}
\newcommand{\Hone}{\mathrm{H}^1}
\newcommand{\DiscSpace}[1]{V_#1}
\newcommand{\DiscSpaceVector}[1]{\boldsymbol{V}_#1}
\newcommand{\Cell}{T}
\newcommand{\triangulation}{\mathcal{T}}
\newcommand{\meshparam}{h}
\newcommand{\norm}[1]{\lVert #1 \rVert }
\newcommand{\seminorm}[1]{\lvert #1 \rvert}
\newcommand{\abs}[1]{\seminorm{#1}}
\newcommand{\Inner}[3][]{\left({#2}\,,\,{#3}\right)\ifthenelse{\equal{#1}{}}{}{_{#1}}}
\newcommand{\InnerApprox}[2]{\Inner[\meshparam]{#1}{#2}}
\newcommand{\Apply}[2]{\left<{#1}\,,\,{#2}\right>}
\newcommand{\restrict}[2]{\left.#1\right\vert_{#2}}

\newcommand{\reducedVolume}{V_r}
\newcommand{\sign}[1]{\operatorname{sign}\left(#1\right)}
\newcommand{\numDiss}{\varepsilon_\meshparam}
\newcommand{\conf}{\mathrm{conf}}
\newcommand{\BC}{B_{\mathrm{c}}}
\theoremstyle{plain}

\theoremstyle{definition}

\newtheorem{Problem}{Problem}
\newcommand{\amdis}{\texttt{AMDiS}}
\newcommand{\dune}{\textsc{Dune}}
\newcommand{\alugrid}{\textsc{Dune-alugrid}}
\newcommand{\curvedgrid}{\textsc{Dune-curvedgrid}}
\newcommand{\functions}{\textsc{Dune-functions}}
\newcommand{\eigen}{Eigen}
\newcommand{\Repulsor}{\texttt{Repulsor}}
\usepackage{mathtools}
\usepackage{ifthen}
\usepackage{subcaption}
\usepackage{pgfplots}
\DeclareUnicodeCharacter{2212}{−}
\usepgfplotslibrary{groupplots,dateplot}
\usetikzlibrary{patterns,shapes.arrows}
\pgfplotsset{compat=newest}
\usepackage{transparent}
\newcommand{\mathdefault}[1]{\displaystyle #1}
\pgfplotsset{
  every axis/.style={
    label style={font=\footnotesize},
    tick label style={font=\tiny},
    title style={font=\small\bfseries},
    legend style={font=\footnotesize},
  }
}

\numberwithin{equation}{section}

\hypersetup{colorlinks=false, pdftitle=Self-avoiding fluid deformable surfaces}

\begin{document}
\definecolor{darkgray176}{RGB}{176,176,176}
\definecolor{forestgreen4416044}{RGB}{44,160,44}
\definecolor{lightgray204}{RGB}{204,204,204}
\definecolor{sienna1408675}{RGB}{140,86,75}
\definecolor{steelblue31119180}{RGB}{31,119,180}
\definecolor{darkcyan27158119}{RGB}{27,158,119}
\definecolor{chocolate217952}{RGB}{217,95,2}
\definecolor{deeppink23141138}{RGB}{231,41,138}
\definecolor{lightslategray117112179}{RGB}{117,112,179}
\definecolor{blue}{HTML}{4477AA}
\definecolor{red}{HTML}{EE6677}
\definecolor{green}{HTML}{228833}
\definecolor{yellow}{HTML}{CCBB44}
\definecolor{cyan}{HTML}{66CCEE}
\definecolor{violet}{HTML}{AA3377}
\definecolor{grey}{HTML}{BBBBBB}
\newcommand{\ColorFB}{green}
\newcommand{\ColorFKin}{red}
\newcommand{\ColorTP}{blue}
\newcommand{\ColorConf}{cyan}
\newcommand{\ColorP}{green}
\newcommand{\ColorNumDiss}{violet}
\newcommand{\ColorDF}{cyan}
\newcommand{\ColorDiss}{yellow}
\newcommand{\ColorPG}{red}
\newcommand{\ColorPhiCluster}{red}
\newcommand{\ColorPhiRep}{green}
\newcommand{\ColorPhiFull}{blue}
\newcommand{\ColorStrong}{green}
\newcommand{\ColorWeak}{violet}

\title{Self-avoiding fluid deformable surfaces}

\author{Maik Porrmann~\orcidlink{0009-0009-5210-8041}}
\email{maik.porrmann@tu-dresden.de}
\affiliation{Institute of Scientific Computing, Technische Universität Dresden, 01062 Dresden, Germany}

\author{Sören Bartels~\orcidlink{0000-0002-8084-5112}}
\email{bartels@math.uni-freiburg.de}
\affiliation{Mathematical Institute, Universität Freiburg, 79104 Freiburg im Breisgau, Germany}

\author{Axel Voigt~\orcidlink{0000-0003-2564-3697}}
\email{axel.voigt@tu-dresden.de}
\affiliation{Institute of Scientific Computing, Technische Universität Dresden, 01062 Dresden, Germany}
\affiliation{Cluster of Excellence Physics of Life (PoL), Technische Universität Dresden, 01062 Dresden, Germany}
\affiliation{Center for Systems Biology Dresden (CSBD), Pfotenhauerstraße 108, 01307 Dresden, Germany}


\begin{abstract}
We propose a numerical method for fluid deformable surfaces governed by surface Stokes flow and Helfrich bending energy under active growth, aiming to model shape evolution of the epithelium in developmental processes. To prevent self-intersections, which commonly arise under large deformations or low enclosed volume to area ratios, we incorporate the nonlocal tangent-point energy to penalize non-embedded configurations. The resulting formulation is discretized using higher order surface finite elements, with a parallelizable assembly strategy for the nonlocal terms. To tailor mesh quality to the geometric evolution, we propose a curvature-adaptive mesh redistribution strategy that improves mesh resolution in regions of high curvature. Numerical examples include the discocyte-to-stomatocyte transition and the inversion of a sphere within a spherical confinement. Both demonstrate the robustness of the method in capturing large deformations, self-avoidance, symmetry-breaking and growth-induced morphology changes.
\end{abstract}

\maketitle

\section{Introduction}
Fluid deformable surfaces are ubiquitous interfaces in biological systems, playing an
essential role in processes from the subcellular to the tissue scale. A prominent example on the larger scale is the epithelium, whose dynamic behavior underlies key developmental processes. Notably, morphogenetic events such as gastrulation can be associated with epithelium folding \cite{davidson2012epithelial}. From a mechanical point of view, fluid deformable surfaces are soft materials exhibiting a solid–fluid duality: while they store elastic energy when stretched or bent, like solid shells, they cannot do so under in-plane shear, a situation
under which they flow as two-dimensional, viscous fluids. For an example of fluid-like tissues in embryonic morphogenesis see \cite{mongera2018fluid}. A common modeling approach for fluid deformable surfaces \cite{Waxman_SAM_1984,PhysRevE.75.041605,Arroyoetal_PRE_2009,torres-sanchezModellingFluidDeformable2019,bachiniDerivationSimulationTwophase2023} is based on the combination of the classical bending model of Helfrich \cite{Helf73} and the modeling of the hydrodynamics of fluid films of Scriven \cite{Scriven_CES_1960}. Omitting inertial effects, the resulting relations correspond to the force and torque balance equations and the constitutive laws postulated in \cite{salbreux2017mechanics} as well as the thin-film limit of the corresponding bulk model equations in \cite{nitschke2019hydrodynamic}. The equations are characterized by a tight coupling between tangential flows and shape changes in the presence of curvature. Any shape change contributes to strains and so must be accompanied by a tangential flow and vice versa, tangential flows on curved surfaces induce shape deformations. This coupling makes curvature a natural element of the governing equations of fluid deformable surfaces, not only with respect to the equilibrium shape but also the dynamics. It thus provides one essential step towards the simulation of developmental processes in biology \cite{al2021active}.

Several numerical approaches have been proposed to solve models for fluid deformable surfaces. Most studies are restricted to an axisymmetric setting \cite{Arroyoetal_PRE_2009,mietke2019self,olshanskii2023equilibrium} or other constraints, e.g. simply connected surfaces \cite{torres-sanchezModellingFluidDeformable2019}. For numerical approaches without these restrictions we refer to \cite{reuther2020numerical,nestlerStabilityRotatingEquilibrium2023,Zhu_Saintillan_Chern_2025,garcke2025parametric}. Analytical results have been obtained for the full set of equations \cite{WaZZ12}, for the Navier-Stokes(-Cahn-Hilliard) model on prescribed evolving surfaces \cite{elliott2024navier}, for the Navier-Stokes model on stationary surfaces \cite{pruss2021navier} and if surface hydrodynamics is neglected for the corresponding Willmore flow, e.g. \cite{kuwert2002gradient,simonett2001willmore} and the references therein. Numerical analysis results going beyond results on stability \cite{garcke2025parametric} so far only exist for special cases, e.g. Stokes flow on stationary surfaces \cite{harderingParametricFiniteelementDiscretization2024,reusken2025analysis} and Willmore flow, see \cite{dziuk2008computational,doi:10.1137/070700231} among others. Building on these results, numerical algorithms for model extensions have been considered. These include constraints on the enclosed volume \cite{Krause2022}, surface two-phase flow problems \cite{bachiniDerivationSimulationTwophase2023,SISCHKA2025118166}, and active geometric forces are considered in \cite{porrmannShapeEvolutionFluid2024}. Higher order bending terms are the subject of \cite{sischkaInfluenceHigherOrder2025} and dynamic responds to area and volume changes has been addressed in \cite{krauseWrinklingFluidDeformable2024}. All of them permit self-intersections even though fluid deformable surfaces cannot pass through themselves.

While unlikely or even impossible to occur if the ratio of enclosed volume to surface area is close to a sphere, already for a reduced volume $\reducedVolume < 0.5$ (which considers this ratio, see Section~\ref{sec::notation} or \cite{seifertShapeTransformationsVesicles1991} for a definition), the local minimizers of the Helfrich energy along the oblate branch show self-intersections \cite{seifertShapeTransformationsVesicles1991}. Therefore, by considering the dynamics in such situations -- like the already mentioned developmental process of gastrulation \cite{GB_2017} -- self-intersections are likely to occur using the proposed algorithms. To avoid them, the evolution must be constrained to remain within the space of \textit{embedded} surfaces. One way to realize this is to include an energy that avoids non-embedded surfaces such as the \textit{tangent-point energy}, originally proposed for knots by \cite{buckSimpleEnergyFunction1995} and analyzed for elastic curves and shells in \cite{strzelecki2012,strzeleckiTangentPointRepulsivePotentials2013, blattENERGYSPACESTANGENT2013}, among others. This energy is infinite for surfaces which are only immersed but not embedded, such that dynamics obeying some energy balance never lead to self-intersections. Due to the non-locality of the resulting term in the equations the computational effort increases significantly. Only recently progress has been made to reduce this effort \cite{bartelsSimpleSchemeApproximation2018,yuRepulsiveCurves2021,yuRepulsiveSurfaces2021,sassenRepulsiveShells2024}. We follow some of these ideas and adapt them to avoid self-intersections in surface finite element methods (SFEM) \cite{dziuk2013finite,nestlerFiniteElementApproach2019} for fluid deformable surfaces. This also requires to deal with the mesh quality of the evolving surface. The challenges are similar to classical geometric flows and established smoothing heuristics, which incorporate appropriate tangential mesh movements, such as the BGN approach \cite{barrettParametricApproximationWillmore2008}, can, in principle, be applied. However, due to the coupling with surface hydrodynamics, shape evolutions happen more rapidly and the additional terms related to surface growth and self-avoidance lead to more complex shapes, in particular shapes with strongly varying curvature. This increases the challenge and requires surface meshes which are not just regular, but adaptive to curvature. We therefore propose a modified version of the BGN approach, which adjusts the mesh to equidistribute the surface curvature and yields better results for the considered examples.

The paper is structured as follows: In Section~\ref{sec::model}, we introduce the notation and the mathematical model for fluid deformable surfaces and focus on self-avoidance, confinement, and active growth. The last two aspects reflect the current understanding that epithelial tissues fold via buckling induced by cell proliferation in confined geometries \cite{trushko2020buckling}. In Section~\ref{sec::numerics}, we describe a SFEM approach to solve the problem. Details are provided for the parallelized treatment of the \textit{tangent-point energy} and the considered curvature-adaptive mesh redistribution strategy. Section~\ref{sec::results} presents computational results which demonstrate the effectiveness of the numerical approach. Finally, in Section~\ref{sec::discussion}, we discuss the results and draw conclusions.

\section{Modeling}
\label{sec::model}
\subsection{Notation}
\label{sec::notation}
We use a similar notation as in \cite{bachiniDerivationSimulationTwophase2023} which is here repeated for convenience.
We consider a time dependent smooth and oriented surface $\S = \S(t)$ without boundary, embedded in $\R^3$ and given via a parametrization $\param : \baseManifold \to \S$ on some base manifold $\baseManifold$.
The enclosed volume is denoted by $\Omega = \Omega(t)$. We define the reduced volume as the scaled quotient of the enclosed volume and the surface area $\reducedVolume= 6 \sqrt{\pi} \abs{\Omega} / \abs{\S}^{3/2} \in (0,1]$, such that for a sphere $\reducedVolume = 1$, see \cite{seifertShapeTransformationsVesicles1991}.
We denote by $\normal$ the outward pointing surface normal, the surface projection is $\P=\Id-\normal\otimes\normal$, with $\Id$ the identity matrix, the shape operator is $\B= -\gradC\normal$, the mean curvature $\H= \tr\B$, and the Gaussian curvature $\K = \frac{1}{2}\left(\H^2-\|\B\|^2\right)$. We consider time-dependent Euclidean-based $ n $-tensor fields in $\restrict{\tangent^n\R^3}{\S}$. We call $\restrict{\tangent^0\R^3}{\S} = \tangent^0\S$ the space of scalar fields, $\restrict{\tangent^1\R^3}{\S}=\restrict{\tangent\R^3}{\S}$ the space of vector fields, and $\restrict{\tangent^2\R^3}{\S} $ the space of 2-tensor fields.
Important subtensor fields are tangential $n$-tensor fields in $\tangent^n\S \le \restrict{\tangent^n\R^3}{\S}$.
Let $p \in \tangent^0\S$ (surface pressure) be a continuously differentiable scalar field, $\flow \in \restrict{\tangent\R^3}{\S}$ (surface velocity) a continuously differentiable $\R^3$-vector field, and $\stress \in \restrict{\tangent\R^3}{\S} \otimes \tangent\S$ (surface rate of deformation tensor) a continuously differentiable $\R^{3\times3}$-tensor field defined on $\S$. We define the (componentwise) surface gradient by $\gradC p = \nabla p^e\P$, $\gradC\flow = \nabla\flow^e\P$ and $\gradC\stress = \nabla\stress^e \P$,
where $p^e$, $\flow^e$ and $\stress^e$ are arbitrary smooth extensions of $p$, $\flow$ and $\stress$ in the normal direction and $\nabla$ is the gradient of the embedding space $\R^3$. The traces of those operators give rise to one notion of divergence operators, which for a vector field $\flow$ and a $2$-tensor field $\stress$ are $\DivC\flow = \tr(\gradC\flow)$ and $\DivC \stress = \tr_{(2,3)}(\gradC\stress)$, where $\tr$ is the trace operator.
However, those divergences are not the adjoints of the gradients, which we will denote by $\divC = -\gradC^*$.
The relations to the covariant derivative $\gradS$ and the covariant divergence $\divS$ on $\S$, read $\gradC p=\gradS p$ and ${\DivC\flow = \divS(\P\flow)-(\flow\cdot\normal)\H}$, in contrast to $\divC \flow = \divS(\P \flow)$, respectively. Lastly we can define the componentwise Laplace operator $\lapC = \divC\circ\gradC$, which agrees with the Laplace-Beltrami operator $\lapS = \divS \circ \gradS$ only for scalar fields, to write the $(2,3)$-contracted Gauss-Weingarten equations as $\H\normal=\lapC \param$ with a common abuse of notation.

\subsection{A self-avoiding Stokes-Helfrich model with active growth}
\label{sec::models}

\subsubsection{Model formulation}
\label{sec::modelformulation}

Following the model derivations in \cite{Arroyoetal_PRE_2009,torres-sanchezModellingFluidDeformable2019,bachiniDerivationSimulationTwophase2023}, among others, we consider the Stokes-Helfrich model for fluid deformable surfaces.
\begin{Problem}[Stokes-Helfrich model for fluid deformable surfaces]
  \label{prob::Stokes-Helfrich}
  At each time $t$ given a parametrization $\param$, find $\vec{u} \in \Hone(\S)^{3}$, $p \in \Hone(\S)$, $\lambda \in \R$, such that
  \begin{align}
    \!\!\!\frac{1}{\Bending}(\lapS \H + \H( \norm{ \B }^{2} - \frac{1}{2}\H^{2} ))\normal
    + \gradS p + p\H\normal + \lambda \normal &= \frac{1}{\Reynolds}\DivC \stress - \friction \vec{u} \\
    \DivC \vec{u}& =0   \\
    \int_{\S} \vec{u} \cdot \normal \,\areaelement{} & =0,\\
    \frac{\partial}{\partial t}(\param) \cdot \normal - \flow \cdot \normal & = 0, \label{eq::meshMovement1}\\
    \H\normal - \lapC \param & = 0, \label{eq::structuralEquation1}
  \end{align}
  where $\stress = \frac{1}{2}\left(\P \gradC \vec{u} + (\P \gradC \vec{u})^T\right)$.
\end{Problem}
Here the first equation describes the balance of viscous and bending forces in the regime of low Reynolds numbers, with the constraints for local inextensibility and conservation of the enclosed volume included as Lagrange-multipliers $p$ and $\lambda$ as prescribed by the second and third equations, respectively. Thereby $\Bending$ is the bending capillary number, $\Reynolds$ the Reynolds number and $\friction$ the friction coefficient. Moreover, eq.~\eqref{eq::meshMovement1} relates the material flow to the change in the surface parametrization, while eq.~\eqref{eq::structuralEquation1} closes the system with a geometric identity. This model describes the dynamics of fluid deformable surfaces as long as the ratio between enclosed volume and surface area is large enough and variations of it have been considered numerically in \cite{torres-sanchezModellingFluidDeformable2019, Krause2022,porrmannShapeEvolutionFluid2024, nestlerStabilityRotatingEquilibrium2023,Zhu_Saintillan_Chern_2025,sischkaInfluenceHigherOrder2025,SISCHKA2025118166}.

In this paper, we extend this model into two directions. First, we add a regularizing energy contribution, that penalizes self-penetration. Second, we use this new capability to consider evolution under confinement and active surface growth. To treat confinement we penalize near contact with a confining surface and active growth is considered similar to \cite{krauseWrinklingFluidDeformable2024}, but on a much larger scale, aiming to trigger not just wrinkling but to model dynamics driven by strong changes in enclosed volume and surface area. Effectively, these extensions simply result in new right hand sides for each of the equations in Problem~\ref{prob::Stokes-Helfrich}.

\begin{Problem}[Self-avoiding Stokes-Helfrich model for fluid deformable surfaces with active growth]
  \label{prob::repulsiveStokesHelfrich}
At each time $t$ given a parametrization $\param$, find $\vec{u} \in \Hone(\S)^{3}$, $p \in \Hone(\S)$, $\lambda \in \R$, such that
  \begin{align}
    \!\!\!\frac{1}{\Bending}(\lapS \H + \H( \norm{ \B }^{2} - \frac{1}{2}\H^{2} ))\normal
    + \gradS p + p\H\normal + \lambda \normal &= \frac{1}{\Reynolds}\DivC \stress - \friction \vec{u} - \frac{\delta \TP + \delta \TP_{\conf}}{\delta \param} \label{eq:SH1}\\
    \DivC \vec{u} &= \delta_{A}
    \label{eq:SH2}\\
    \int_{\S} \vec{u} \cdot \normal \,\areaelement{} &= \delta_{V},
    \label{eq:SH3}\\
    \frac{\partial}{\partial t}(\param) \cdot \normal - \flow \cdot \normal & = 0,
\label{eq::meshMovement}\\
    \H\normal - \lapC \param & = 0,
      \label{eq::structuralEquation}
  \end{align}
 where the actions of the repulsive forces on some vector field are given by $\frac{\delta \TP}{\delta \param}$ and $\frac{\delta \TP_{\conf}}{\delta \param}$, see below, and $\delta_{A}$ and $\delta_V$ denote the area and volume growth rates, respectively.
\end{Problem}

\begin{figure}[ht]
  \def\svgwidth{0.9\linewidth}
  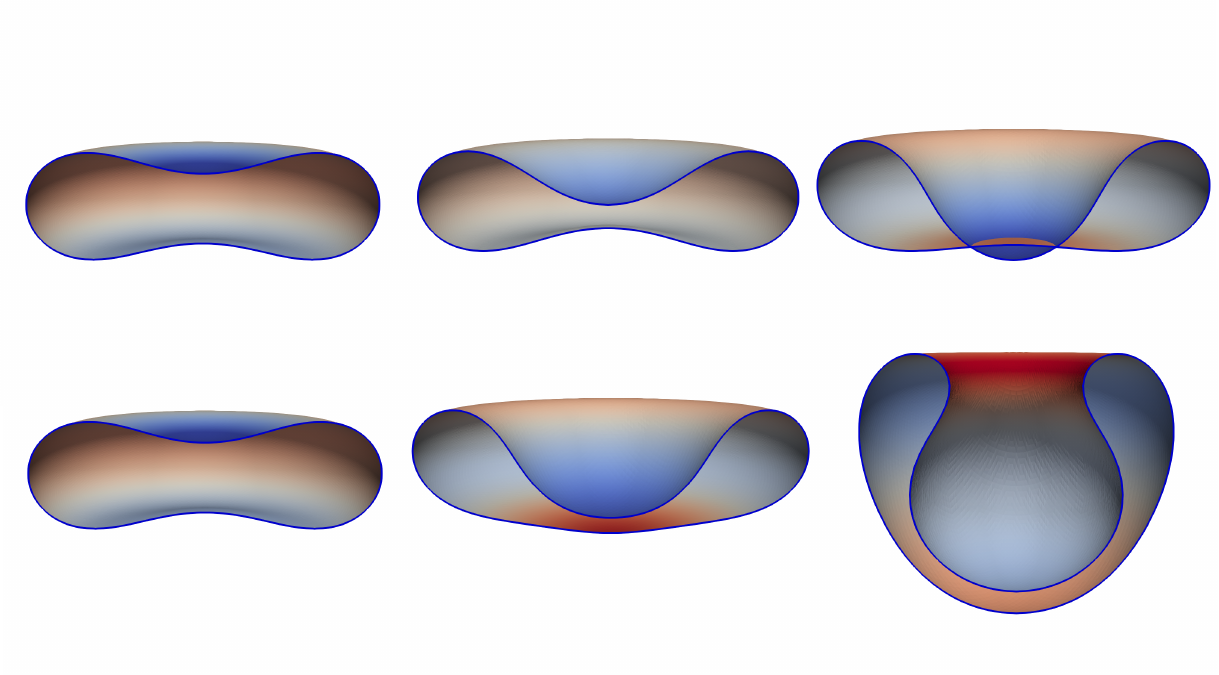
\caption{Cuts through an evolving surface under localized area growth. Top: Model without repulsive forces leads to self-intersections. Bottom: Repulsive forces prevent self-intersection. The color coding shows the normal component of the velocity $\flow \cdot \normal$, red indicating movement outwards, blue indicating movement inwards. The considered parameters are the same in both simulations and are described in Sec. \ref{sec::results}. }\label{fig::self-intersection}
\end{figure}

Before we discuss these new terms in detail we show their impact on the evolution of a growing surface in Fig.~\ref{fig::self-intersection}. In the first case active area growth without repulsive forces leads to an unphysical self-intersection, in the second case with repulsive forces such states can be prevented.

\subsubsection{Self-avoidance}
\label{sec::model-self-avoidance}
Real objects disallow self-intersections, i.e. when deformed they stay embedded. While for small deformations this is practically guaranteed when minimizing a bending energy, for large deformations, as well as for models of fluid deformable surfaces under constraints like conservation of the enclosed volume, self-intersecting states can naturally arise or even become minimizers. The Stokes-Helfrich model is such a case, where for $\reducedVolume \lessapprox 0.5$ \cite{seifertShapeTransformationsVesicles1991} the minimizers of the so-called oblate branch, discocytes reminiscend of red blood cells, display self-intersections.
To incorporate self-avoidance into our model we consider the \textit{tangent-point energy}. It's definition relies on the \textit{tangent-point radius}
\begin{equation}
  R(\pos,\posy) \coloneqq  \frac{\norm{ \pos - \posy }^{2}}{2\seminorm{ \normal(\pos)\cdot (\pos-\posy) } },
\end{equation}
which is the radius of the smallest sphere that is tangent to $\S$ at $\pos$ and intersects $\S$ at $\posy$.
\begin{figure}
  \def\svgwidth{0.7\linewidth}
  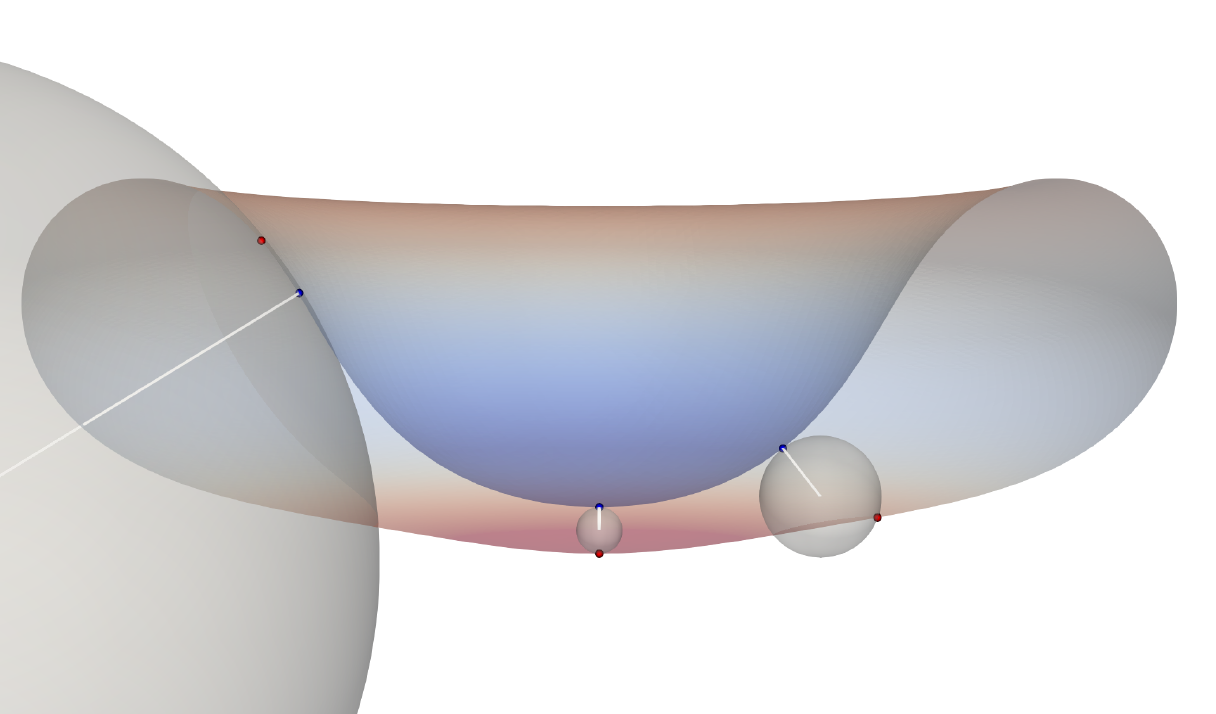
  \caption{Tangent-point radius $R(\pos, \posy)$ (white lines) for three point pairs. The spheres are tangent to the blue points $\pos_i$ and intersect the surface at the red points $\posy_i$. Note the crucial feature, that for two geodesically close points, here $(\pos_0,\posy_0)$, the radius is very large. The radius is smallest for $(\pos_1, \posy_1)$. The considered surface is the same as in Figure~\ref{fig::self-intersection}.}
\end{figure}
The tangent-point energy $\TP(\param)$ is then defined as
  \begin{equation}
    \label{eq::TP}
    \TP(\param)\coloneqq \rho \frac{2^{-\TPexponent}}{\TPexponent}\int_{\S} \int_{\S} \left(\frac{1 }{R(\pos,\posy)}\right)^\TPexponent  \areaelement{\posy}  \areaelement{\pos} = \frac{\rho}{\TPexponent}\int_{\S} \int_{\S} \frac{\seminorm{\normal(\pos)\cdot (\pos-\posy) }^{\TPexponent} }{\norm{ \pos - \posy }^{2\TPexponent}}  \areaelement{\posy}  \areaelement{\pos},
  \end{equation}
with $\TPexponent > 4$. The central feature of this energy is that, restricting ourselves to sufficiently regular closed surfaces, it is finite if and only if the surface has no self-intersections.

Inspired by the work in \cite{yuRepulsiveCurves2021, yuRepulsiveSurfaces2021, sassenRepulsiveShells2024, bartelsSimulatingSelfAvoidingIsometric2022}, who demonstrated the viability of (linear finite element) schemes involving this non-local energy, we include it as an additional potential energy in our model, effectively forming an elastic barrier around the surface. When forced to close contact situations, this barrier stores energy and releases it when the external force is removed. Note, that we also could follow \cite{sassenRepulsiveShells2024} and add a dissipative term derived from the tangent-point energy to our system, effectively assigning a high energy cost to paths that lead to self contact. However, the physical justification of such a term is out of scope of this paper. In any case, given the large gradients of $\TP$ near self contact, we expect a regularizing effect from the potential energy rather than elastic behavior. Naturally, this behavior depends on the choice of the parameters $\rho$ and $\TPexponent$. For analytical reasons, we have $\TPexponent>4$, such that we set $\TPexponent=6$ from here on. The parameter $\rho$ is in principle free, and should be chosen such that in a state far from self-intersections, the tangent-point energy is small compared to other energies in the system. In our simulations, we typically set $\rho = 10^{-4}$.

\subsubsection{Confinement}
\label{sec::confinement}
With slight modifications the tangent-point energy $\TP(\param)$ can also be used to model evolution within a confinement. To consider (self-)avoidance with a fixed confining surface $\S_{\conf}$ we define the \textit{confinement energy} $\TP_{\conf}(\param)$ as
  \begin{equation}
    \label{eq::TPConf}
    \TP_{\conf}(\param)\coloneqq \rho_{\conf}\frac{2^{-\TPexponent}}{\TPexponent}\int_{\S} \int_{\S_{\conf}} \left(\frac{1 }{R(\pos,\posy)}\right)^\TPexponent \, \areaelement{\posy}  \, \areaelement{\pos} .
  \end{equation}
Again we consider $\TPexponent = 6$ throughout and $\rho_\conf = 10^{-4}$. Within the envisioned application of gastrulation the confinement is a protective elastic shell \cite{bailles2019genetic,munster2019attachment}. We simply consider the confinement to be a sphere in all simulations. However, the above choice of $\TP_{\conf}$ also allows for non-smooth confinements, since the normal, and therefore the derivatives of the parametrization are only evaluated on the confined surface, but not on the confinement itself.

\subsubsection{Active growth}
\label{sec::model-growth}
As already introduced in \cite{krauseWrinklingFluidDeformable2024} to consider wrinkling we use active growth terms $\delta_A$ for the surface area as well as $\delta_V$ for the enclosed volume. Eq. \eqref{eq:SH2} is considered in analogy to free boundary problems for Stokes flow to model growth of biological tissue \cite{king2021free}. Within the context of surface Stokes flow it has also been used in \cite{toshniwal2021isogeometric}. $\delta_A \neq 0$ violates the conservation of mass on the surface. It effectively models area growth, which might result from microscopic phenomena, such as cell proliferation or variations in thickness of the surface as a consequence of a constant cellular volume. Previous approaches have considered $\delta_A = const$ in space. However, there is no need for this restriction and in view of the mentioned modeling aspects a spatially varying $\delta_A$ seems even more realistic. In general this might result from couplings with other fields, e.g. concentrations of nutrients or a variable for the thickness of the surface. Here we simply consider $\delta_A(\pos)$. Considering $\delta_V \neq 0$ describes influx or outflux through the surface and essentially a coupling with the bulk phase. As demonstrated experimentally in \cite{krauseWrinklingFluidDeformable2024}
if appropriately scaled both $\delta_A$ (if considered to be constant) and $\delta_V$ lead to a change in the reduced volume $V_r$ and essentially cause the same behavior. We will therefore only concentrate on $\delta_A$ in the simulations and set $\delta_V = 0$ from here on.
\section{Numerical approach}
\label{sec::numerics}
We follow the numerical scheme detailed in \cite{Krause2022,porrmannShapeEvolutionFluid2024}. In particular, the approach considers a surface finite element method (SFEM) \cite{dziuk2013finite,nestlerFiniteElementApproach2019} in space and a semi-implicit finite difference method in time within an Arbitrary Lagrangian-Eulerian (ALE) approach \cite{elliottALEESFEMSolving2012,Krause2022,SAUER2025118331}, primarily Lagrangian in normal direction and Eulerian in tangential direction.

\subsection{Spatial discretization}
We consider a discrete $k$-th order approximation $\S_h^k$ of $\S$, given by a curved mesh $\triangulation$, with $h$ the size of the mesh elements, i.e. the longest edge in $\triangulation$. We consider each geometrical quantity like the normal vector $\normal_h$, the shape operator $\B_h$, the Gaussian curvature $\K_h$, and the inner products $(\cdot , \cdot)_h$ with respect the $\S_h^k$. In the following we will drop the index $k$ and only write $\S_h$. We define the discrete function spaces for scalar functions by $\DiscSpace{k}(\S_h)=\{ \psi \in C^0(\S_h) \,\vert \, \restrict{\psi}{\Cell}\in \mathcal{P}_{k}({\Cell}) \,\forall \Cell \in \triangulation\}$ and for vector fields by $\DiscSpaceVector{k}(\S_h)=[\DiscSpace{k}(\S_h)]^3$. Within these definitions $\Cell$ is the $k$-th order mesh element and $\mathcal{P}_{k}$ are the polynomials of order $k$. We consider $\flow_h,\param\in \DiscSpaceVector{3}(\S_h)$,  $\H_h\in \DiscSpace{3}(\S_h)$, and $p_h\in \DiscSpace{2}(\S_h)$, which leads to an isogeometric setting for the velocity and a $\mathcal{P}_{3}-\mathcal{P}_{2}$ Taylor-Hood element for the Stokes-like system on the surface. For a stationary surface this setting is analyzed in \cite{harderingParametricFiniteelementDiscretization2024}.
We discretize in time using a semi-implicit time stepping scheme with constant step size $\tau$. In each time step $t^n$ we solve the surface Stokes-like equations and the mesh movement together. To this end we define a discrete surface update variable $\paramUpdate^{n}=\param^{n}-\param^{n-1}$, which is considered as unknown instead of the surface parametrization $\param^{n}$.
Treating all geometric terms with respect to $\param^{n-1}$ leads to a linear system, which reads:
\begin{Problem}[Fully discrete Problem]
  Given a surface described by a parametrization $\param^{n-1}$ and a timestep $\tau$, find $(\flow_h^n,\pressure_h^n,\H_h^n,\paramUpdate^n, \lambda)\in[\DiscSpaceVector{3}\times \DiscSpace{2}\times \DiscSpace{3} \times \DiscSpaceVector{3}\times \R](\S_h^{n-1})$ such that:
  \begin{align}
    \!\!\!\!- \frac{1}{\Bending} \InnerApprox{ \gradS \H_h^n }{\gradS(\flowTest_h\cdot\normal_h^{n-1})}
     + \frac{1}{\Bending} \InnerApprox{\H_h^n \mathcal{C} }{\flowTest_h \cdot \normal_h^{n-1}}&\nonumber
    \\
    + \frac{1}{\Reynolds} \InnerApprox{
      \stress(\flow_h^{n})}{\gradC\flowTest_h} + \friction \InnerApprox{\flow_h^n}{\flowTest_h}\nonumber
      &
    \\
    + \InnerApprox{p_h^n}{\DivC\flowTest_h}
    + \lambda \InnerApprox{\normal_h^{n-1}}{\flowTest_h}&= - \Apply{\frac{\delta \TP + \delta\TP_{\conf}}{\delta \param}}{\flowTest_h}
      \\
      \InnerApprox{\DivC\flow_h^{n}}{\pressureTest_h} &= \InnerApprox{\delta_A}{\pressureTest_h}
      \\
      \InnerApprox{\flow_h^{n} \cdot \normal_h^{n-1}}{\mu} &= \InnerApprox{\delta_V}{\mu}
      \\
      \InnerApprox{\paramUpdate^n_h \cdot \normal^{n-1}_h}{\HTest_h} - \tau\InnerApprox{\flow^n_h\cdot \normal^{n-1}_h}{\HTest_h} &= 0
      \\
      \label{eq::discStructuralEquation}
      \InnerApprox{\H_h^n \normal_h^{n-1}}{\paramUpdateTest_h}
      + \InnerApprox{\gradC \paramUpdate^n}{\gradC \paramUpdateTest_h}
      &=
      - \InnerApprox{\gradC \param^{n-1}}{\gradC \paramUpdateTest_h}
    \end{align}
    for all $(\flowTest_h, \pressureTest_h, \HTest_h, \paramUpdateTest_h, \mu )\in[\DiscSpaceVector{3}\times \DiscSpace{2}\times \DiscSpace{3}\times\DiscSpaceVector{3}\times \R](\S_h^{n-1})$.
  \end{Problem}
    The new parametrization is then given by \begin{equation}
    \param^n=\param^{n-1} + \paramUpdate^n,
    \end{equation}
    which has been exploited in \eqref{eq::discStructuralEquation} using the BGN method \cite{barrettParametricApproximationWillmore2008}. Above, $\InnerApprox{\cdot}{\cdot}$ denotes the $\Ltwo$ inner product on the discrete surface $\S_h$ and  $\mathcal{C}=\norm{\B_h^{n-1}}^2-\frac{1}{2} (\H^{n-1})^2$ is the squared norm of the deviatoric part of the shape operator. We compute $\mathcal{C}$ directly from the grid, but one can also use the solution $\H^{n-1}_h$ of the previous timestep. For practical stability however, only the implicit treatment of $\H_h^n$ is important.
\subsection{Assembly of the nonlocal operator}
\label{sec::AssemblyTP}
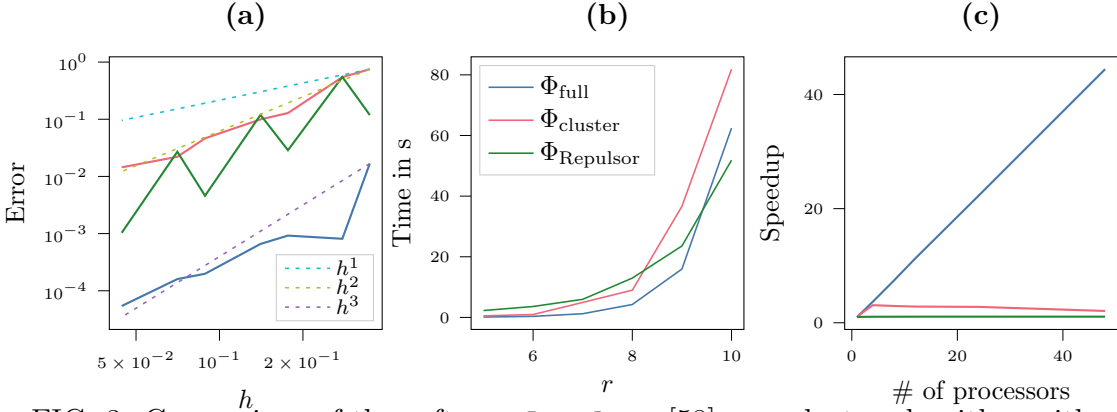
\begin{figure}[htbp]
\centering
    \input{TP_h_Theta10.tex}
    \input{TP_h_Theta10_runtime_v2.tex}
    \input{TP_h_Theta5_r10_speedup_v2.tex}\vspace{-1.8em}
  \caption{Comparison of the software \Repulsor{} \cite{sassenRepulsiveShells2024}, our cluster algorithm with $\theta=10\meshparam$ and our full integration. All computations were done with 48 shared-memory cores for a dumbbell-like shape, for which we computed the reference value $\TP_{\mathrm{ref}}=1.60987907$ for the convergence study with our full integration on refinement level $r_{\mathrm{ref}} = 15$.
  (a)~Approximation error in $\TP$, note, that the irregularity in \Repulsor{}'s accuracy is due to our bisection refinement. (b)~Runtime for $10$ evaluations of the differential $\frac{\delta \TP}{\delta \param}$, using $48$ threads. Note, that the near constant runtime for \Repulsor{} for small meshes is probably due to their adaptive refinement strategy. (c)~Speedup for $r=10$. The legend in (b) is valid for all three plots. Moreover, in (b)~and (c), we have added selected measurements for a shape as in the final plot of Fig.~\ref{fig::no_sphere_confinement}, which is much closer to self-contact. Note, that these values for \Repulsor{} can be improved by remeshing algorithms, as recommended in \cite{sassenRepulsiveShells2024}, but this renders the discrete space incompatible to our cubic setup. We also want to stress, that the results shown here will not carry over to other hardware or threading scenarios.}
  \label{fig::TPComp}
\end{figure}

One of the major challenges is the assembly of the nonlocal terms in $\frac{\delta \TP}{\delta \param}$ (and respectively in $\frac{\delta \TP_{\conf}}{\delta \param}$). First, we note that the differential can be straightforwardly calculated with the formulas from \cite{nitschke_observer-invariant_2022, nitschkeTensorialTimeDerivatives2023} for the deformation derivative in direction $\paramUpdate \in \restrict{\tangent \R^3}{\S}$, defined by
\[\deformation{\paramUpdate} f \coloneqq \restrict{\frac{d}{d\varepsilon}}{\varepsilon = 0}\!\!\!\!\!\!(\restrict{f}{\param + \varepsilon \paramUpdate}) \] for a quantity $f \in \tangent^0\S$ defined on a moving surface $\S = \param (\baseManifold)$.
We obtain the equation
\begin{align}
  \!\!\!\!\Apply{\frac{\delta \TP}{\delta \param}}{\paramUpdate} = & \, \rho \int_{\S} \frac{1}{\TPexponent} \DivC(\paramUpdate) \int _{\S}K(\pos,\posy)\, \areaelement{\posy} + \frac{1}{\TPexponent}\deformation{\paramUpdate} \int _{\S}K(\pos,\posy) \, \areaelement{\pos} \nonumber\\
  = & \rho\int_{\S} \frac{1}{\TPexponent}\DivC(\paramUpdate)(\pos) \int_{\S} K(\pos,\posy) \, \areaelement{\posy} + \int _{S} \frac{1}{\TPexponent} \DivC(\paramUpdate)(\posy) K(\pos,\posy) \, \areaelement{\posy} \, \areaelement{\pos} \nonumber\\
   &+ \rho \int_{\S}  \int_{\S} \frac{1}{\TPexponent}\deformation{\paramUpdate}K(\pos,\posy)   \, \areaelement{\posy} \, \areaelement{\pos},
  \label{eq::dTP}
\end{align}
where
$$
  K(\pos,\posy) \coloneqq \left(\frac{1}{R(\pos,\posy)}\right)^\TPexponent
$$
and
\begin{equation*}
  \begin{split}
    \deformation{\paramUpdate} K(\pos,\posy)  =& \frac{\deformation{\paramUpdate} \seminorm{\normal(\pos) \cdot (\pos - \posy)}^{\TPexponent}}{\norm{ \pos  - \posy}^{2\TPexponent}} + \seminorm{\normal(\pos) \cdot (\pos - \posy)}^{\TPexponent}\deformation{\paramUpdate}\frac{ 1}{\norm{ \pos  - \posy}^{2\TPexponent}}\\
    =& \TPexponent  \frac{\seminorm{ \normal(\pos) \cdot (\pos - \posy) } ^{\TPexponent}}{\norm{ \pos - \posy }^{2\TPexponent}}\frac{\normal(\pos) \cdot (\paramUpdate(\pos) - \paramUpdate(\posy)) -\normal(\pos) \cdot \nabla_{C}\paramUpdate(\pos) \cdot (\pos - \posy)}{\normal(\pos)\cdot(\pos - \posy)} \\
    &- 2\TPexponent \frac{\seminorm{ \normal(\pos) \cdot (\pos - \posy) } ^{\TPexponent}}{\norm{ \pos - \posy }^{2\TPexponent}}\frac{(\pos - \posy) \cdot (\paramUpdate(\pos) - \paramUpdate(\posy))}{\norm{ \pos - \posy }^{2}}.
  \end{split}
\end{equation*}
The nonlocal nature of eq. \eqref{eq::dTP} has two consequences for computational schemes. First, the double integral leads to a computational cost quadratic in the number of quadrature points. We can alleviate this by either approximating eq. \eqref{eq::dTP} with a Barnes-Hut \cite{barnesHierarchicalLogForcecalculation1986} like cluster strategy as done in \cite{yuRepulsiveSurfaces2021} and improved in \cite{sassenRepulsiveShells2024} for flat triangles ($\S_h^1$), or implement the assembly in a parallel manner. Since transferring a curved mesh (e.g. $\S_h^3$) to a flat and back introduces additional errors, which seemed to dominate the resulting gradients of the tangent-point energy in our experiments, we cannot simply use the software \Repulsor{} form \cite{sassenRepulsiveShells2024} and therefore explore an alternative implementation strategy. We remark that for both the clustering and the parallel strategy, in our situation it seems mandatory to cache the evaluation of positions, surface elements and normals at all quadrature points, as the computation of those quantities is much more expensive for curved mesh elements ($\S_h^3$) than for flat ones. To be beneficial, data structures for those caches need to be as contiguous in memory as possible and the algorithms need to iterate those caches synchronously in memory order as our experiments suggest that the cost of fetching data from memory dominates the actual computation, at least in a parallelized setting. Since the interplay of clustering and parallelization is complex, we compare in Fig.~\ref{fig::TPComp} the software from \cite{sassenRepulsiveShells2024}, our own Barnes-Hut like implementation, and a full integration scheme, which only aims to optimize parallelization. Our experiments suggest that the optimal choice of method highly depends on the number of mesh elements, the shape of the surface and the hardware used. In particular, when computing on JUWELS supercomputer at J\"{u}lich Supercomputing Centre with $48$ shared memory threads but relatively low single core performance, the clustering approaches outperformed the full integration only for relatively large meshes, see Fig.~\ref{fig::TPComp}(b), while the approximation accuracy of the full integration scheme substantially benefits from the higher resolution of the surface, see Fig.~\ref{fig::TPComp}(a).

The second disadvantage of the nonlocal structure concerns time discretization. Naturally one would like to treat these terms in an implicit manner, but this necessarily leads to a dense linear system to solve, with the computational cost of solving exploding accordingly. For the problem of minimizing the tangent-point energy $\TP(\param)$, \cite{yuRepulsiveCurves2021,yuRepulsiveSurfaces2021} have developed a framework of fractional gradient flows with special solvers dedicated to this. Such modification of the dynamics is certainly acceptable in the considered applications in \cite{yuRepulsiveCurves2021,yuRepulsiveSurfaces2021}, but in a physical setting we cannot simply choose a different gradient flow. Hence, we have to restrict ourselves to explicit treatment of eq. \eqref{eq::dTP}.

We briefly comment on the concrete formulation used in our assembling routine, as this discussion is so far missing in literature.
As long as the surface stays away from self-intersections, we can exploit the symmetry in the domains of integration to obtain a formulation more suitable to finite element methods:
\begin{align}
  \label{eq::dTP_reordered}
  \Apply{\frac{\delta \TP}{\delta \param}}{\paramUpdate} &= \rho\int_{\S} f_{K}(\pos)\DivC(\paramUpdate)(\pos) + \vec{f}(\pos)\cdot \paramUpdate(\pos) - \normal(\pos) \cdot \nabla_{C}\paramUpdate(\pos) \cdot \vec{f}_{\normal}(\pos)\, \areaelement{\pos},
\\\text{where} &\nonumber\\
{f}_{K}(\pos) &= \frac{1}{\TPexponent} \int_{\S} \left(K(\pos,\posy) + K(\posy,\pos)\right) \, \areaelement{\posy}  \\
\vec{f}_{\normal}(\pos) &= \int_{\S} \frac{\seminorm{ \normal(\pos) \cdot (\pos - \posy) } ^{\TPexponent -1}}{\norm{ \pos - \posy }^{2\TPexponent+2}} \sign{\normal(\pos)\cdot(\pos - \posy)} \norm{ \pos - \posy }^{2}(\pos - \posy)\,\areaelement{\posy}  \\
\vec{f}(\pos) &= \int_{\S} \frac{\seminorm{ \normal(\pos) \cdot (\pos - \posy) } ^{\TPexponent -1}}{\norm{ \pos - \posy }^{2\TPexponent+2}} \sign{\normal(\pos)\cdot(\pos - \posy)} \norm{ \pos - \posy }^{2}  \normal(\pos)
\nonumber \\
&\qquad\quad- \frac{\seminorm{ \normal(\posy) \cdot (\posy - \pos) } ^{\TPexponent -1}}{\norm{ \posy - \pos }^{2\TPexponent+2}}\sign{\normal(\posy)\cdot(
  \posy - \pos)} \norm{ \posy - \pos }^{2}  \normal(\posy) \nonumber \\
&\qquad\quad -2 \frac{\seminorm{ \normal(\pos) \cdot (\pos - \posy) } ^{\TPexponent -1}}{\norm{ \pos - \posy }^{2\TPexponent+2}}\seminorm{ \normal(\pos) \cdot (\pos - \posy) }\,(\pos - \posy) \nonumber \\
& \qquad \quad +2 \frac{\seminorm{ \normal(\posy) \cdot (\posy - \pos) } ^{\TPexponent -1}}{\norm{ \posy - \pos }^{2\TPexponent+2}} \seminorm{ \normal(\posy) \cdot (\posy - \pos) }\,(\posy - \pos) \, \areaelement{\posy} .
\end{align}

We highlight the advantage of using eq. \eqref{eq::dTP_reordered}. Directly implementing eq. \eqref{eq::dTP}, with the inner integrals depending on the test function $\paramUpdate$, leads to a quadratic number of write operations when transferring element local assemblies into the system vector. However, in eq. \eqref{eq::dTP_reordered} the test function $Y$ is only evaluated in the outer integral, allowing the interpretation of $f_K, \vec{f}_{\normal}$ and $\vec{f}$ as functions and hence straightforward inclusion in standard finite element assembly routines.
Moreover, the form of eq. \eqref{eq::dTP_reordered} enables a clear separation of clustering and parallelization. Following the ideas of \cite{yuRepulsiveCurves2021} we only aim to cluster all inner integral terms, effectively using precomputed averages over a patch of mesh elements, whenever a certain admissibility condition is fulfilled. For details, we refer to \cite{yuRepulsiveSurfaces2021}. Since the inner integrals $f_K, \vec{f}_{\normal}$ and $\vec{f}$ are independent of the direction of variation, we do not need to back propagate derivatives as done in \cite{yuRepulsiveSurfaces2021,sassenRepulsiveShells2024}. On the other hand, as mentioned in \cite{yuRepulsiveSurfaces2021}, only clustering the inner integral also enables a top level parallelization approach, where the assembly of the outer integral is done in parallel, and, provided a synchronization of the potentially very large caches, even allows domain decomposition methods on distributed grids. However, in our case we employed a shared memory parallelization to avoid costly synchronization.

In order to validate our modified approach, we briefly compare our approximation schemes in terms of accuracy of approximation and runtime with the software \Repulsor{} from \cite{sassenRepulsiveShells2024}, which uses a piecewise linear approximation both for the surface and for its displacement vectors, in a setting suitable for our application. Firstly note, that such a comparison between a linear triangle approximation and a cubic triangle approximation is inherently difficult. To accommodate this gap to some extent, we created a linear mesh to be used by \Repulsor{} from our cubic mesh by dividing each cubic triangle into $9$ linear triangles. In this manner, we obtain discrete meshes for both approaches, which have exactly the same amount of degrees of freedom, located at the same points in space. All computations were done on a mesh with sphere topology and $24\times 2^r$ cubic triangles, where $r$ is the number of global bisection refinements. As mentioned, our cluster implementation only reduces the computation of the inner integrals in contrast to \Repulsor{}, which clusters both integrals. Both approaches depend on a \textit{separation parameter}, denoted by $\theta$, which controls how eagerly the algorithm clusters. In our case, when measuring the accuracy of approximation of $\TP$, we observe close to quadratic convergence, if we scale $\theta \in  \mathcal{O}(\meshparam)$, see Fig.~\ref{fig::TPComp}(a), even though we can only expect linear convergence. A similar effect can be observed for \Repulsor. For the full integration scheme we observed a convergence close to order $3$, as expected by the order of surface representation. As expected, our full integration scheme (using order $3$ and $4$ quadrature rules for the outer and inner integral, respectively) is more accurate than both approximations. In Fig.~\ref{fig::TPComp}(b) the computing time and in Fig.~\ref{fig::TPComp}(c) the parallel speedup are shown. Due to the good parallelization speedup the full integration scheme also achieves acceptable runtime, at least for coarser shapes. Our key observation is, that the parallelized full integration scheme can compete with \Repulsor's superior cluster algorithm for finer meshes than we anticipated. As expected however, on a single-core machine \Repulsor{} outperforms the full integration scheme much earlier. We remark, that for both our clustering algorithm as well as \Repulsor{} absolute runtime and speedup depend on the discrete surface, while the full integration scheme only depends on the number of triangles and the quadrature scheme used, which allows us to balance the times needed for assembling of the non-local contributions and solving the algebraic system.
\subsection{Curvature-adaptive mesh redistribution}
  \label{sec::tangentialMovement}
  The surface mesh moving in space has a priori no guarantee to remain regular. Additionally, when updating the surface, the error hinges on the accuracy of the normal representation. It is hence desirable to use a finer mesh at areas of large curvature. To remedy this, we introduce an artificial mesh movement $f_{\mathcal{T}} \in \tangent\S$ in tangential direction. By the reparametrization invariance of our continuous problem, this has no effect on the shape of the continuous solution. Note, that by following \cite{barrettParametricApproximationWillmore2008}, we already have some tangential motion induced by eq. \eqref{eq::discStructuralEquation}, for which ad hoc forces between vertices have been proposed and recently shown to converge \cite{bai2025convergenceanalysisbarrettgarckenurnbergmethod}.

  Mesh optimization by geometrical vertex movement rather than topological refinement has some history for flat problems, most notably \textit{moving mesh methods}, see \cite{huangMovingMeshPartial1994}. On surfaces, where the subject is even more critical, only few approaches are known. In particular, \cite{crestelMovingMeshMethods2015, kolasinskiSurfaceMovingMesh2020} apply the ideas of \cite{huangMovingMeshPartial1994} to surfaces. However, in \cite{crestelMovingMeshMethods2015} a mesh in the parameter domain is optimized and then mapped onto the surface via a fixed parametrization, contrary to our moving surface which mandates a changing parametrization. The approach in \cite{kolasinskiSurfaceMovingMesh2020} is closer to our requirements in that they optimize the surface mesh, even depending on curvature, but base their derivations on vertices of linear mesh elements. Our experiments suggest, that a straightforward application of such a force to higher order curved elements leads to mesh tangling.

  A second line of surface mesh distribution approaches has been initiated by \cite{DeTurckElliot}, who used the so-called DeTurck trick to reparametrize the mean curvature flow and surface diffusion by a harmonic flow field. The notable difference to a moving mesh method is that this approach does not directly depend on the mesh, but tries to construct a close to conformal map from a fixed reference manifold onto the surface. This allows for convergence analysis as well as for a precise notion of flow. Quite recently, similar approaches have been integrated into the BGN method \cite{barrettParametricApproximationWillmore2008} and other schemes, see \cite{huEvolvingFiniteElement2022, duanNewArtificialTangential2024,duanMeshPreservingEnergyStableParametric2025}. However, those approaches are not trivial to implement as they mix energies on a reference grid and on the moving surface, such that including them into our scheme is beyond the scope of this paper, and more importantly, none of the above consider curvature dependent mesh optimization.

  Leaving the transfer of the more general ideas in \cite{huangMovingMeshPartial1994,kolasinskiSurfaceMovingMesh2020,DeTurckElliot} to future work, we here propose a very simple approach to define a continuous force with the desired properties and include it without the need for additional variables.
  Our goal is to equidistribute the total curvature, that is $\int_{\Cell}^{}\norm{\B}^2 \areaelement{}$ should be similar for all $ \Cell \in \triangulation$. A simple approach to achieve this is to set
  \begin{equation}
    \label{eq::meshMovementFormula}
  \P \restrict{\paramUpdate_h}{\Cell}(\pos ) = \gradS \left(\norm{\B}^2(\pos)g_{\Cell}(\pos) \right) \eqqcolon f_{\triangulation}(\pos),
\end{equation}
where $g_{\Cell} = \sqrt{\det(J_{\Cell}^\top J_{\Cell})}$ is the Gramian determinant of the mapping from the reference element $\hat{\Cell}$ to $\Cell$, in other words, the Riemannian metric density when interpreting the mesh $\triangulation$ as an atlas. It is easy to see, that if $f_{\triangulation} \equiv \vec{0}$, we have the desired equidistribution, since for all $\Cell$ we have:
\begin{align}
  g_\Cell &= C/\norm{\B}^2 \label{eq::equidistributionPointwise}\\
  \int_{\Cell} \norm{\B}^2dS &= \int_{\hat{\Cell}} \norm{\B}^2 g_{\Cell} d\hat{\pos} = \int_{\hat{\Cell}} C d\hat{\pos}, \label{eq::equidistribution}
\end{align}
where $C\seminorm{\hat{\Cell}}$ is the average of $\norm{\B}^2$ per element in the surface triangulation.
Less obvious is the argument, that eq. \eqref{eq::meshMovementFormula} leads towards such a state. Naively, we can argue that the gradient points in the directions of high curvature to area element ratio. It therefore moves area in that direction, which is desired. Instead of a proof we demonstrate its practical applicability.
Note, that $\norm{\B}^2$ is not strictly positive and even very small positive value would lead to a very large $g$ in near flat areas. In turn, $g$ should also not become arbitrarily small. We therefore replace $\norm{\B}^2$ in the formulas above by a clamped version
\[ \BC(\pos) \coloneqq \min\left(L, \max\left(M,\norm{\B}^2(\pos)\right)\right),\]
with $L \in \R$ a small and $M \in \R$ a large parameter.

\begin{figure}[htbp]
  \def\svgwidth{0.9\linewidth}
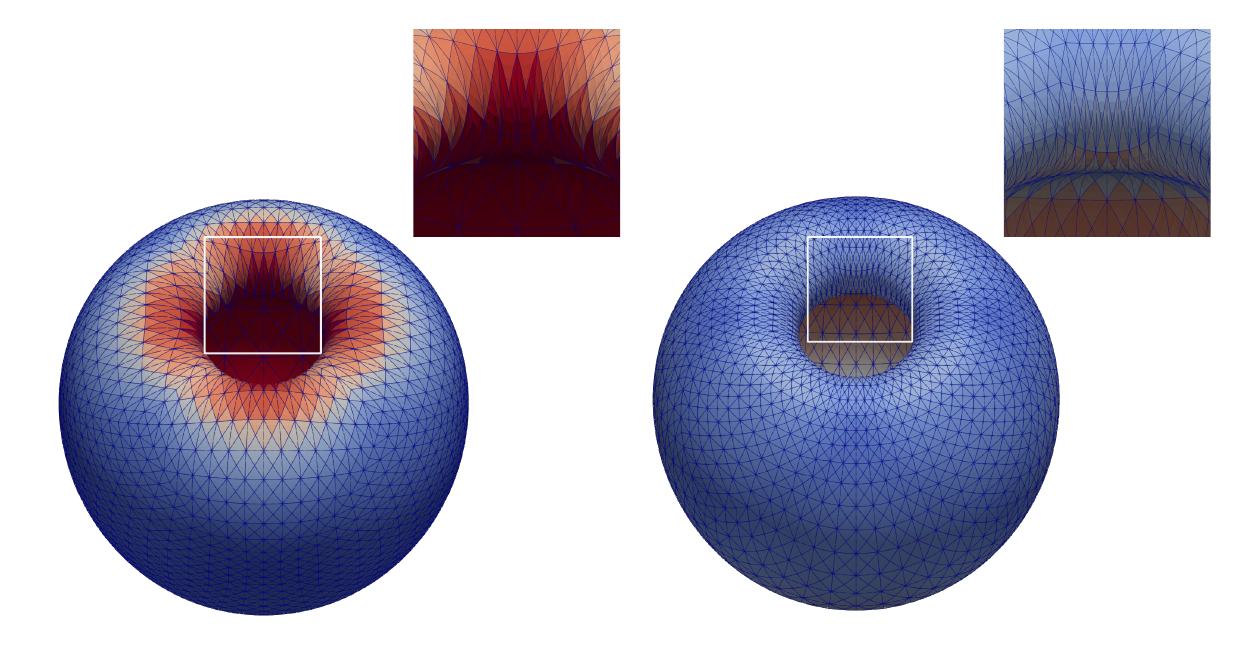
\caption{Effect of the mesh redistribution approach for the discocyte to stomatocyte transition at refinement level $r = 8$. Parameters are $\tilde{\epsilon} = 10^{-6}$, $M=10$, and $\epsilon=1$ (left) and $\epsilon = 100$ (right). The color indicates $\int_{\Cell}\BC d\S$, i.e. the quantity we aim to equidistribute. While the overall triangle quality also improves, our approach cannot fully prevent the distortion of mesh elements. Note that the closeup views in the respecitive top rights are taken from a different angle.}
\label{fig::meshRedistribution}
\end{figure}

\begin{figure}[htbp]
  \input{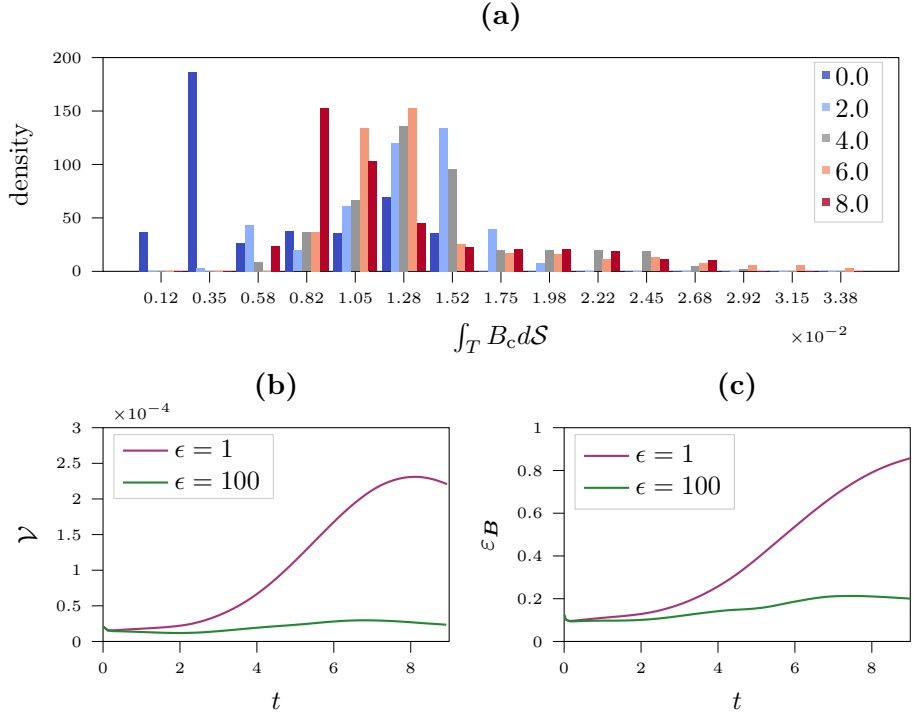}
  \caption{Effect of the proposed mesh redistribution approach. (a) A histogramm of the distribution of $\int_{\Cell} \BC d\S$ at different times and $\epsilon =100$. (b) The variance of the same quantity over time for two different $\epsilon$. (c) Quantification of the violation of the pointwise equidistribution condition over time. The corresponding simulation is discussed in Sec. \ref{sec::oblateToStomatocyte}. }
  \label{fig::meshStats}
\end{figure}

Recall, that equation \eqref{eq::discStructuralEquation}, the discretized weak version of \eqref{eq::structuralEquation}, leaves the tangential part of $\lapC \paramUpdate$ undetermined. Hence we can add any small tangential force such as $\epsilon f_\triangulation$ to move the mesh. In practice, we add $\InnerApprox{\tau \epsilon \gradS \tilde{f}_h}{\paramUpdateTest_h}$ to \eqref{eq::discStructuralEquation}, where
\begin{align}
  \tilde{f}_h = (\tilde{\epsilon} + \lapS) \mathbb{P}_{\DiscSpace{k}}\left(\BC(\pos) g(\pos)\right)
\end{align}
is solved as a separate auxiliary equation to obtain a smoothed version of the clamped $f_\triangulation$. Above, $\epsilon \in \R$ scales the speed of redistribution, $\tilde{\epsilon} \in \R$ is small, and $\mathbb{P}_{\DiscSpace{k}}$ is a projection onto $\DiscSpace{k}$ that averages the discontinuous values of $\BC$ and $\sqrt{\det{F^\top F}}$. For our experiments, we set $\tilde{\epsilon} = 10^{-6}$, $L=1$, $M= 10$ and only varied $\epsilon$. The choices of $L$ and $M$ effectively constrain the triangle sizes to a deliberately narrow range.
Note that the clamping between $L$ and $M$ means that the algorithm will not optimize towards \eqref{eq::equidistribution}, but the clamped version of it, and that we do not claim that this algorithm converges for a stationary surface nor can we prove that it avoids mesh tangling.
In Fig.~\ref{fig::meshRedistribution} we show the effect of our approach for a very strongly deformed surface for small and large $\epsilon$. The enhancement of mesh quality is clearly visible.
Additionally, we quantify the effect of our scheme on the evolution discussed in Sec.~\ref{sec::oblateToStomatocyte} in Fig.~\ref{fig::meshStats}. Specifically, we evaluate the sample variance of the clamped curvature per mesh element, that is
\[
\mathcal{V} \coloneqq \operatorname{s^2}\left(\int_{\Cell}\BC\areaelement{}\right),
\]over $\Cell \in \triangulation$ as well as the violation of \eqref{eq::equidistributionPointwise} for the clamped curvature, i.e.
\[
  \varepsilon_{\B}(\pos) \coloneqq \norm{\BC g_{\Cell}(\pos) - \frac{\int_{\S}\BC \areaelement{}}{\seminorm{\hat{\Cell}}\seminorm{\triangulation}}}_{\Lone}.
\]
While both demonstrate that the approach is not perfect, as $\varepsilon_{\B} > 0$ and $\mathcal{V} > 0$ throughout, Fig.~\ref{fig::meshStats} demonstrates the strong improvement compared to the standard BGN approach, which would be close to the case $\epsilon = 1$.

\subsubsection{Software}
We have realized the described discretization within the finite element toolbox \amdis{}  \cite{VeyVoigt2007AMDiS,AmdisOld}, available at \cite{AMDiS:2.10}, which is a high level module in the \dune{} \cite{Dune2.10} ecosystem. For an overview of \dune{} we refer to \cite{duneBook}. Moreover, we use the modules \alugrid{} \cite{dune-alugrid} to manage a grid with sphere topology, \curvedgrid{} \cite{dune-curvedGrid} to map this grid onto the surface by the parametrization $\param_h$. This parametrization, as well as all other discrete functions and their bases are represented by means of \functions{} \cite{dune-functions}. We use \eigen{}  \cite{eigenweb} as a linear algebra backend, and the direct solver PARDISO as well as the multithreading framework TBB from Intel's oneAPI \cite{intel2025oneapi}. Finally, all 3D visualizations have been created with ParaView \cite{Paraview}.
\section{Simulation results}
\label{sec::results}
We demonstrate the applicability and robustness of our method with two examples addressing the phenomena of sphere inversion. Again guided by the application to gastrulation, which establishes the first symmetry-break in embryogenesis and involves nonuniform cell growth and confinement, we examine the ability of local area growth to induce sphere inversion. This is considered in two scenarios, first by a discocytoe to stomatocyte transition, and second by spherical shape within a sphere confinement. In both cases we break the symmetry, in the first case by considering local growth on one side of the discocyte and in the second by slightly flattening the spherical shape on one side. The last part of this section is dedicated to validation of the computational scheme. Due to the computational complexity of the tangent-point energy, we omit a full convergence study, but focus on the numerical dissipation as an indicator for error.

\subsection{Stomatocyte shapes induced by localized growth}
\label{sec::oblateToStomatocyte}
The first experiment considers the discocyte to stomatocyte transition. Our general setup is to start with an oblate ellipsoid with $\reducedVolume\approx 0.7$, which relaxes to a biconcave minimizer of the Helfrich energy $\Helfrich = \frac{1}{\Bending}\int_\S \H^2 \areaelement{}$ without self-intersections. After some time we gradually add localized area growth up to a total strength of $1$ area unit per time unit. This is achieved by setting the growth rate as $\areagrowth(t,\pos) = c(t)g(\pos)$, where the time-dependent cut-off function is defined by
\[c(t) = \frac{1}{2}\left(1 + \tanh\left(\frac{1}{\epsilon_{\mathrm{g}}}\left(t - t_0\right)\right)\right) \in (0,1),\]
and the spatial localization satisfies the normalization condition
 \begin{equation}
  \int_{\S} g(\pos) \areaelement{} = 1. \label{eq::growthStrength}
 \end{equation}
The localized profile $g(\pos)$ is initialized as a normalized Gaussian centered at $\pos_0$ via
 \begin{align}
  \hat{g}(\pos) &= \mathbb{P}_{\DiscSpace{k}} \frac{1}{\sigma \sqrt{2\pi}}\exp\left(-\frac{1}{2}\frac{\norm{\pos - \pos_0}^2}{\sigma^2}\right),\\
  g(\pos) &= \hat{g}(\pos)\frac{1}{\int_S \hat{g}\areaelement{}}.
\end{align}
For all experiments shown here, we use $t_0 = 0.1$, $\epsilon_{\mathrm{g}} = 0.01$ to obtain a smooth but rapid increase of growth and $\sigma=0.25$ and $\pos_0 = (0,0,1)^T$.
For simplicity, we use the time dependence only to allow for a smooth simulation setup and correct the values of $g$ only to obey \eqref{eq::growthStrength} but not against mesh movement.

\begin{figure}[htbp]
\centering
  \def\svgwidth{0.9\linewidth}
  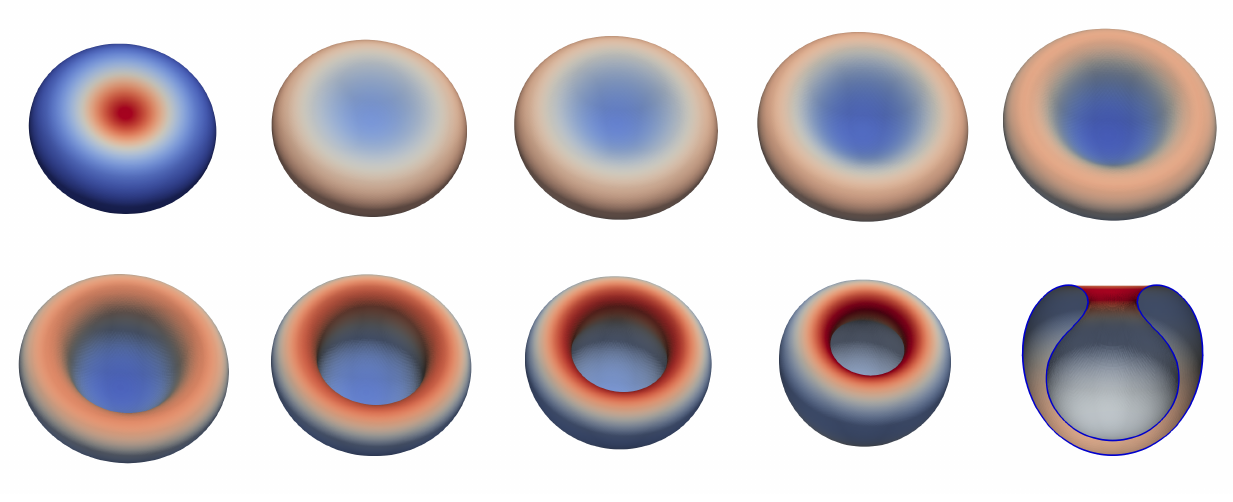
\caption{Discocyte to stomatocyte transition at refinement level $r = 8$ for parameters $\Bending = 10$, $\Reynolds = 1$, and $\friction = 0.1$. Localized growth is defined at the top, $\pos_0=(0,0,1)^T$. The color indicates the shape change density $\flow \cdot \normal$, except for the first image, where it indicates the growth rate. Red corresponds to positive values, blue to negative values. We start from an ellipsoid with $\reducedVolume=0.7$. At $t=9$ we show a cut through the surface, as in Fig. \ref{fig::self-intersection}. The area to volume ratio at this final state is $\reducedVolume \approx 0.315$.}
\label{fig::oblateToSto}
\end{figure}
\begin{figure}[htbp]
\centering
  \def\svgwidth{0.9\linewidth}
  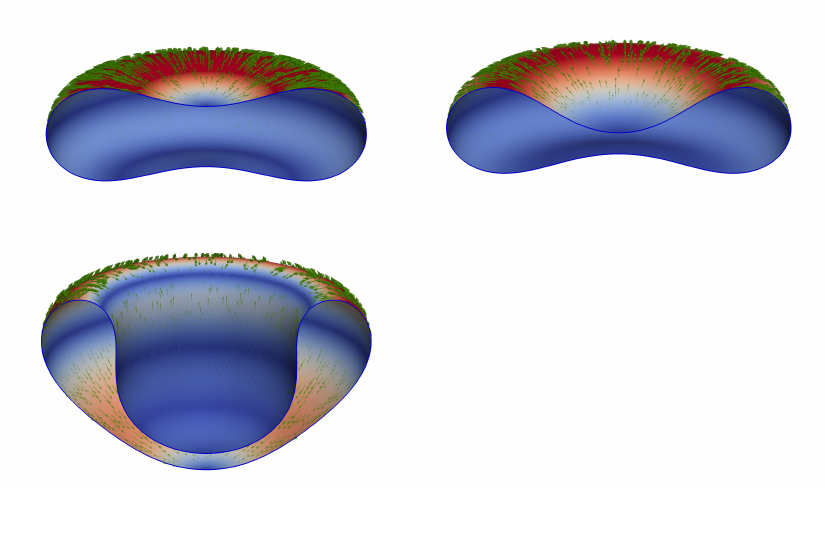
\caption{Cut through selected timesteps of the evolution shown in Fig.~\ref{fig::oblateToSto}. The surface color indicates the norm of the tangential flow $\norm{\P \flow}_{\Ltwo}$ (blue indicates low and red high values), and $\P \flow$ is also depicted by the green arrows. For $t = 6.0$ a full view is provided as well.}
\label{fig::oblateToSto_glyphs}
\end{figure}

A prototypical evolution is shown in Fig.~\ref{fig::oblateToSto}. The shape evolves along a path close to the minimizers of the respective reduced volumes, but with stronger concavity on site of growth (top) than opposite to it. Once the biconcave minimizers would show self-intersections, the evolution tends towards a stomatocyte ($t \approx 4.0$). The shape remains essentially rotational symmetric with respect to $z$-axis. However the up-down symmetry of the initial state is broken by the localized growth. While not violating the symmetry the growth induced force is strong enough to drive the evolution out of the local minimizers and is sufficient to induce the transition to a stomatocyte. For the same evolution, Fig.~\ref{fig::oblateToSto_glyphs} visualizes the emerging tangential flows at selected time instances. The emerging patterns in the tangential flow result from the complex interplay of growth and curvature. At the beginning $t = 0.2$ and $t = 2.0$ the tangential flow induced by growth is visible. This changes at $t = 4.0$ where inward bending in part changes the direction of flow and almost annihilates $\P \flow$. This corresponds to the transition towards a stomatocyte. At later times with a developed stomatocyte, shown at $t = 6.0$, more complex flow patterns emerge. However, like the shape, the tangential flow also remains essentially symmetric.

\subsection{Sphere inversion in a spherical confinement}
\label{sec::confinedSpheres}
We now consider the effect of $\TP_{\conf}$, with the confining surface $\S_\conf$ chosen as a sphere of radius $1.1$. As initial condition we consider a sphere of radius $1$. Both spheres are concentric. While in the previous example localized growth was able to break the up-down symmetry of the discocyte, for the increased symmetry of the sphere this is no longer the case. We therefore slightly flatten one side of the sphere to break the symmetry already in the initial condition.
Specifically, in spherical coordinates the initial surface is given by
\[
\left(1-\delta_\S \exp\left(-\frac{\theta^2}{\epsilon_\S^2}\right), \theta, \varphi\right)^T, \quad \theta \in [0,\pi], \, \varphi \in [0,2\pi],
\]
where $\theta$ is the azimuthal angle and $\varphi$ is the polar angle. We set $\delta_\S = 0.01$ and $\epsilon_\S = 0.1$.

We again consider localized growth, as described in the previous section. But we vary its location and investigate whether localized growth can overcome the asymmetry in the initial shape.

\begin{figure}[htbp]
\centering
  \def\svgwidth{0.9\linewidth}
  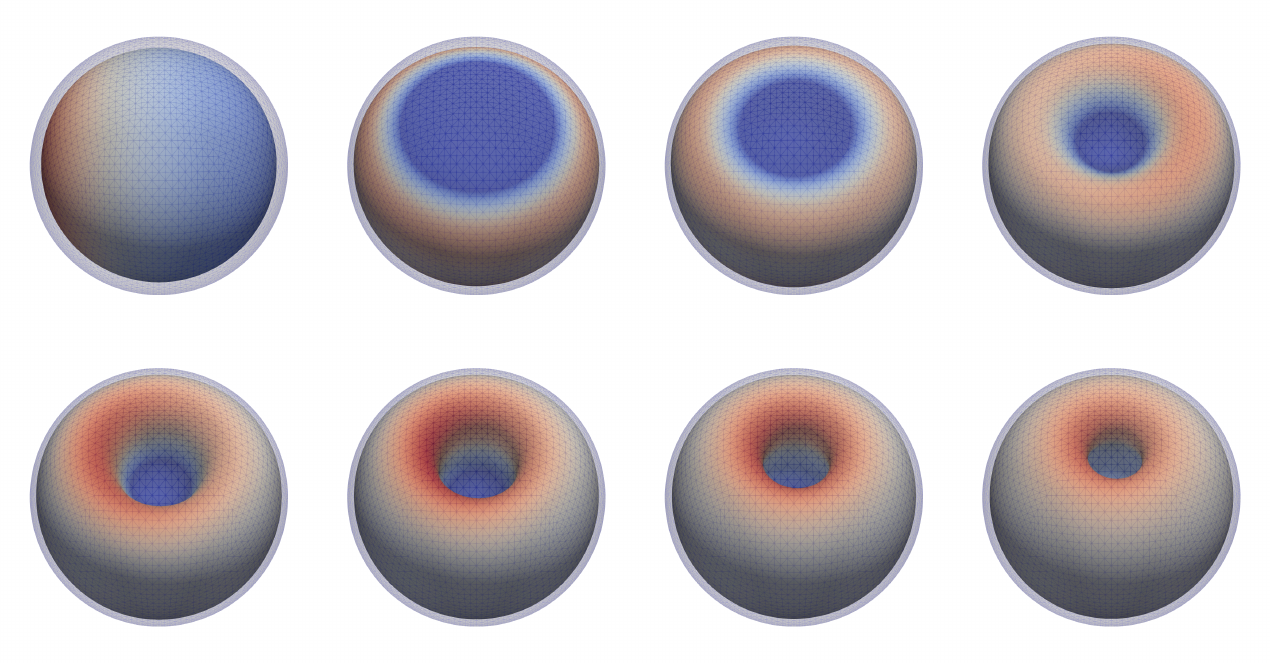
\caption{Sphere inversion in spherical confinement at refinement level $r = 9$ for parameters $\Bending = 1$, $\Reynolds = 1$, and $\friction = 0.1$. The initial shape is slightly flattened (barely visible) at the top. Localized growth is defined at the side, $\pos_0=(0,1,0)^T$. The color indicates the shape change density $\flow \cdot \normal$, except for the first image, where it indicates the growth rate. Red corresponds to positive values, blue to negative values. In addition the confinement is shown. At $t=0$, we have $\reducedVolume \approx 0.99998$, while at $t=3.0$ we have $\reducedVolume \approx 0.7323$.}
\label{fig::no_sphere_confinement}
\end{figure}

In particular, we simulated growth located at the flattened pole, $\pos_0 = (0,0,1)^T$, opposite to it, $\pos_0 = (0,0,-1)$, and at the side, $\pos_0 = (0,1,0)$. Interestingly, the symmetry break in initial shape dominates the effect of localized growth such that all scenarios led to the same phenomena, namely indentation and sphere inversion at the previously flattened location. We only show one of these solutions, the one with $\pos_0 = (0,1,0)$, in Fig.~\ref{fig::no_sphere_confinement}.
As a result of the confinement the sphere inverts and similar to the previous example leads to a stomatocyte shape. We observe, that the flattening, rather than growth, selects the location of indentation. Independent of the growth location, the shape evolves inwards at the top, while evolving outwards elsewhere towards a limit defined by the \textit{tangent-point energy} at the early stage of the evolution ($0.3 < t < 1.0$). When this limit is reached, essentially only the top and inner part further evolve, completing the inversion ($1.5 < t < 3.0$). A slight asymmetry in the shape change density is visible at $t=1.5$. It is a result of the localized growth at the side. This asymmetry leads to a slightly tilted symmetry axis of the final shape.

\subsection{Energy rates and dissipation}
\begin{figure}[htbp]
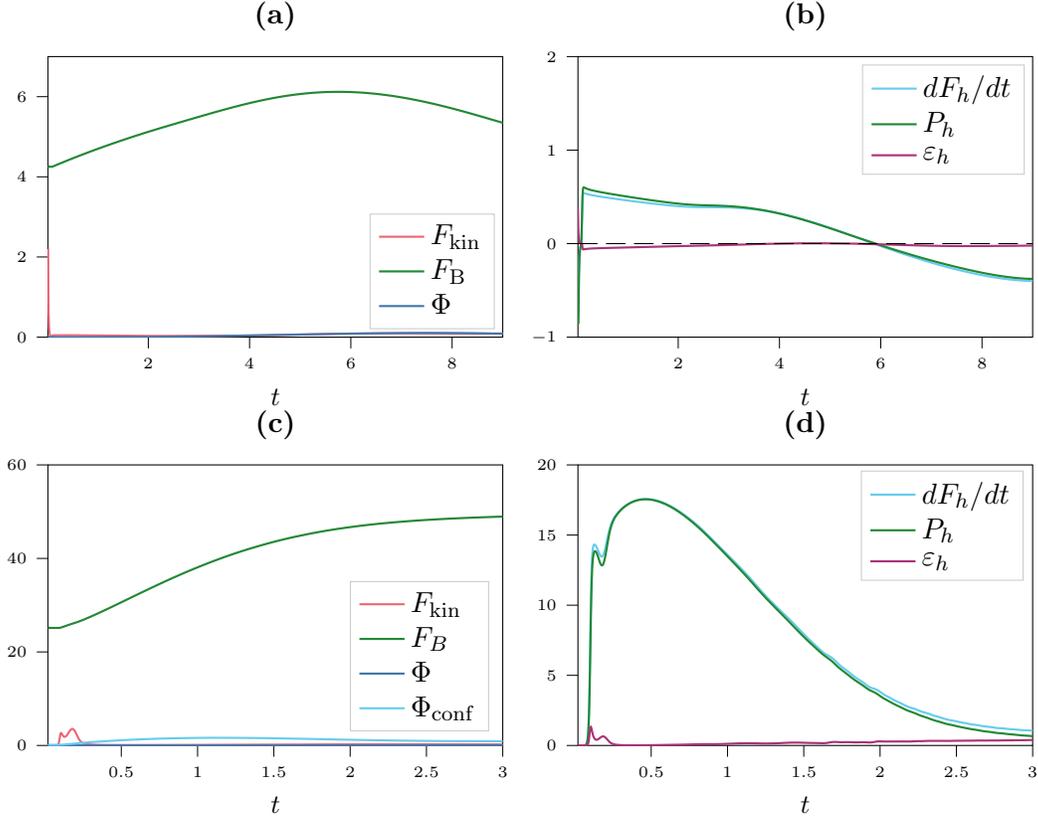

  \centering
  \input{oblateToSto_energies.tex}
  \input{oblateToSto_dissipation.tex}
\input{obstacleNoSphere_energies.tex}
\input{obstacleNoSphere_dissipation.tex}
\vspace*{-1.2em}
\caption{Energy and dissipation over time for the examples in Sec. \ref{sec::oblateToStomatocyte} and \ref{sec::confinedSpheres}. (a) and (b) correspond to the evolution depicted in Fig.~\ref{fig::oblateToSto}, (c) and (d) correspond to the evolution depicted in Fig.~\ref{fig::no_sphere_confinement}.}
\label{fig::energies}
\end{figure}

We briefly analyze the previously discussed evolutions in terms of energy rate and dissipation aiming to verify both model and numerics, and to show key moments in the evolution. To this end, we consider the evolution of various energies as well as the energy rates and dissipation for the examples in the previous sections. For the last we consider \textit{numerical dissipation}, that is, the violation of the discrete energy balance.
Defining the potential energy rate as the time derivative of the Helfrich energy and the tangent-point energies by
\begin{equation}
  \frac{dF}{dt} = \frac{d(\Helfrich + \TP + \TP_{\conf})}{dt}
\end{equation}
and the power emitted by the system in form of viscous and frictional dissipation as well as power input through growth as
\begin{equation}
  P(t) = \int_{S(t)} \frac{1}{\Reynolds}\norm{\stress}^2 + \friction \norm{\flow}^2 - \pressure \DivC(\flow)\areaelement{},
\end{equation}
the continuous problem satisfies the energy balance $\frac{dF}{dt} = -P$ along the trajectories of solutions of Problem~\ref{prob::repulsiveStokesHelfrich}. In the discrete setting, however, numerical errors generally lead to $\frac{dF_h}{dt} + P_h\neq 0$. We consider
\begin{equation}
\numDiss(t_n) \coloneqq \frac{F_h(t_{n+1}) - F_h(t_n)}{t_{n+1} - t_n} + P_h(t_n) ,
\end{equation}
as an indicator of the numerical error.
The results are shown in Fig.~\ref{fig::energies}. The oblate to stomatocyte transition from  Sec.~\ref{sec::oblateToStomatocyte} is considered in Fig.~\ref{fig::energies}(a) and (b). The Helfrich energy $\Helfrich$ is significantly larger than the kinetic energy $F_{\mathrm{kin}}\coloneqq \frac{1}{2}\norm{\flow}_{\Ltwo}^2$, which justifies considering the surface Stokes equations rather than surface Navier-Stokes equations, and also significantly larger than the tangent-point energy $\Phi$, which justifies the regularization approach. Moreover, we can identify different stages in evolution. Once the growth has started, the potential energy is increasing due to power input via active growth. From $t \approx 5.86$ ($\reducedVolume \approx 0.3995)$ on, however, the potential energy is decreasing, while the growth mechanism still leads to a power input, which allows for high viscous dissipation. The evolution of the discrete potential energy rates, the discrete power and $\numDiss$ are shown in Fig.~\ref{fig::energies}(b). The decrease of potential energy is associated with negative values of the discrete potential energy rates and the discrete power. Overall the error remains small.

For the sphere inversion in a spherical confinement from  Sec.~\ref{sec::confinedSpheres} the results are shown in Fig.~\ref{fig::energies}(c) and (d).
Again the Helfrich energy $\Helfrich$ is significantly larger than the kinetic energy $F_{\mathrm{kin}}$ and the {\textit tangent-point energies} $\Phi$ and $\Phi_{\mathrm{conf}}$. The kinetic energy shows a small peak at early stages, which is associated with the initialisation of the indentation. Also a slight increase of $\Phi_{\mathrm{conf}}$ is visible up to $t= 1.0$. Afterwards it remains approximately constant and even decreases towards the evolution of the stomatocyte for $t > 1.5$. In contrast to the oblate to stomatocyte transition, the Helfrich energy always increases. Considering the discrete potential energy rates, the discrete power we see the influence of the peak in kinetic energy at the beginning and the decrease of both over time but remaining positive. Also for this problem the error remains small.
Without showing the plots here, we remark that the energy and dissipation evolutions for growth localized at a different site, as discussed above, show no substantial difference.
\section{Discussion}
\label{sec::discussion}
We have extended the Stokes-Helfrich model for fluid deformable surfaces to prevent both self-intersection of the surface and intersection with a confining surface. We detailed the numerical implementation of the non-local energy contributions and compared and related our approach to an advanced clustering algorithm. Additionally, we proposed a simple extension of the BGN method, that moves grid points tangentially to equidistribute the total curvature. The resulting scheme allows for simulations of surfaces with high area to volume ratios or low reduced volumes $V_r$. To account for that and explore the rich dynamics for such surfaces we have considered local area growth and simulated the transitions of a discocyte to a stomatocyte and a sphere in a spherical confinement which also evolved to a stomatocyte.

Both examples are relevant in the biological context of gastrulation which in the current understanding is induced by cell proliferation (local area growth) in confined geometries \cite{trushko2020buckling}. Both examples are characterized by symmetry breaking and large shape deformations, which are induced by localized growth. In particular, localized area growth can drive the surface along paths close to the minimizers of the respective reduced volume regime. For the considered discocyte  shape it also provides a mechanism to select the location of indentation. With confinement the situation changes. Due to the high symmetry of the considered spherical shapes, localized growth alone seems not able to break symmetry. We therefore slightly flattened the initial shape at one position. Interestingly, this sets the position of the invagination independent of the position of localized growth. Within the context of gastrulation we thus confirm the two expected mechanism of cell proliferation and confinement and in addition propose that surface inhomogeneities (the flattened part) determine the location of invagination, which does not need to agree with the region of growth. In any case, further exploration of the parameter space and comparison to measurements from biology is needed to confirm the model and its applicability in this context.

While the approach enables self-avoidance in models for fluid deformable surfaces and therefore allows to simulate evolutions with large deformations for low reduced volumes, the approach also has several limitations, which give rise to future research. In particular we lack a full convergence study, which is prohibitively expensive due to the computational complexity of the \textrm{tangent-point energy}. Additionally, the mesh distribution approach remains ad hoc and does not prevent distorted mesh elements. Therefore a combination of the approaches in \cite{DeTurckElliot,huEvolvingFiniteElement2022, duanNewArtificialTangential2024, duanMeshPreservingEnergyStableParametric2025} may be considered in the future.
\begin{acknowledgments}
The authors gratefully acknowledge the Gauss Centre for Supercomputing e.V. (www.gauss-centre.eu) for funding this project by providing computing time on the GCS Supercomputer JUWELS \cite{JUWELS} at Jülich Supercomputing Centre (JSC). Moreover, we would like to thank Henrik Schumacher for discussions on the assembly of the tangent-point energy and help with the inclusion of \Repulsor{} into our system.

This work was supported by DFG via research unit FOR 3013 (grant 417223351 to SB and AV) and in part by the Swedish Research Council under grant no. 2021-06594 to the Institut Mittag-Leffler in Djursholm, Sweden.
\end{acknowledgments}

\bibliographystyle{plain}
\bibliography{lit}

\end{document}

%% file: SelfIntersection.pdf_tex
\begingroup%
  \makeatletter%
  \providecommand\color[2][]{%
    \errmessage{(Inkscape) Color is used for the text in Inkscape, but the package 'color.sty' is not loaded}%
    \renewcommand\color[2][]{}%
  }%
  \providecommand\transparent[1]{%
    \errmessage{(Inkscape) Transparency is used (non-zero) for the text in Inkscape, but the package 'transparent.sty' is not loaded}%
    \renewcommand\transparent[1]{}%
  }%
  \providecommand\rotatebox[2]{#2}%
  \newcommand*\fsize{\dimexpr\f@size pt\relax}%
  \newcommand*\lineheight[1]{\fontsize{\fsize}{#1\fsize}\selectfont}%
  \ifx\svgwidth\undefined%
    \setlength{\unitlength}{584.64882678bp}%
    \ifx\svgscale\undefined%
      \relax%
    \else%
      \setlength{\unitlength}{\unitlength * \real{\svgscale}}%
    \fi%
  \else%
    \setlength{\unitlength}{\svgwidth}%
  \fi%
  \global\let\svgwidth\undefined%
  \global\let\svgscale\undefined%
  \makeatother%
  \begin{picture}(1,0.55359827)%
    \lineheight{1}%
    \setlength\tabcolsep{0pt}%
    \put(0,0){\includegraphics[width=\unitlength,page=1]{SelfIntersection.pdf}}%
    \put(0.1322382,0.31122993){\color[rgb]{0,0,0}\makebox(0,0)[lt]{\lineheight{1.25}\smash{\begin{tabular}[t]{l}$t=0.2$\end{tabular}}}}%
    \put(0.46096911,0.31122993){\makebox(0,0)[lt]{\lineheight{1.25}\smash{\begin{tabular}[t]{l}$t=2.0$\end{tabular}}}}%
    \put(0.79369862,0.31122993){\makebox(0,0)[lt]{\lineheight{1.25}\smash{\begin{tabular}[t]{l}$t=4.0$\end{tabular}}}}%
    \put(0.1322382,0.00896483){\makebox(0,0)[lt]{\lineheight{1.25}\smash{\begin{tabular}[t]{l}$t=0.2$\end{tabular}}}}%
    \put(0.46096912,0.00896483){\makebox(0,0)[lt]{\lineheight{1.25}\smash{\begin{tabular}[t]{l}$t=4.0$\end{tabular}}}}%
    \put(0.79369862,0.00896483){\makebox(0,0)[lt]{\lineheight{1.25}\smash{\begin{tabular}[t]{l}$t=8.0$\end{tabular}}}}%
  \end{picture}%
\endgroup%

%% file: tangentPointRadius.pdf_tex
\begingroup%
  \makeatletter%
  \providecommand\color[2][]{%
    \errmessage{(Inkscape) Color is used for the text in Inkscape, but the package 'color.sty' is not loaded}%
    \renewcommand\color[2][]{}%
  }%
  \providecommand\transparent[1]{%
    \errmessage{(Inkscape) Transparency is used (non-zero) for the text in Inkscape, but the package 'transparent.sty' is not loaded}%
    \renewcommand\transparent[1]{}%
  }%
  \providecommand\rotatebox[2]{#2}%
  \newcommand*\fsize{\dimexpr\f@size pt\relax}%
  \newcommand*\lineheight[1]{\fontsize{\fsize}{#1\fsize}\selectfont}%
  \ifx\svgwidth\undefined%
    \setlength{\unitlength}{583.20001557bp}%
    \ifx\svgscale\undefined%
      \relax%
    \else%
      \setlength{\unitlength}{\unitlength * \real{\svgscale}}%
    \fi%
  \else%
    \setlength{\unitlength}{\svgwidth}%
  \fi%
  \global\let\svgwidth\undefined%
  \global\let\svgscale\undefined%
  \makeatother%
  \begin{picture}(1,0.58697917)%
    \lineheight{1}%
    \setlength\tabcolsep{0pt}%
    \put(0,0){\includegraphics[width=\unitlength,page=1]{tangentPointRadius.pdf}}%
    \put(0.61127599,0.23107697){\color[rgb]{0,0,0}\makebox(0,0)[lt]{\lineheight{1.25}\smash{\begin{tabular}[t]{l}$x_2$\\\end{tabular}}}}%
    \put(0.72771092,0.14139947){\color[rgb]{0,0,0}\makebox(0,0)[lt]{\lineheight{1.25}\smash{\begin{tabular}[t]{l}$y_2$\\\end{tabular}}}}%
    \put(0.25494491,0.34543144){\color[rgb]{0,0,0}\makebox(0,0)[lt]{\lineheight{1.25}\smash{\begin{tabular}[t]{l}$x_0$\\\end{tabular}}}}%
    \put(0.22577804,0.38854436){\color[rgb]{0,0,0}\makebox(0,0)[lt]{\lineheight{1.25}\smash{\begin{tabular}[t]{l}$y_0$\\\end{tabular}}}}%
    \put(0.49719493,0.18254821){\color[rgb]{0,0,0}\makebox(0,0)[lt]{\lineheight{1.25}\smash{\begin{tabular}[t]{l}$x_1$\\\end{tabular}}}}%
    \put(0.49950796,0.10806338){\color[rgb]{0,0,0}\makebox(0,0)[lt]{\lineheight{1.25}\smash{\begin{tabular}[t]{l}$y_1$\\\end{tabular}}}}%
  \end{picture}%
\endgroup%

%% file: TP_h_Theta10.tex
\begin{tikzpicture}
\definecolor{darkgray176}{RGB}{176,176,176}
\definecolor{darkorange25512714}{RGB}{255,127,14}
\definecolor{darkturquoise23190207}{RGB}{23,190,207}
\definecolor{forestgreen4416044}{RGB}{44,160,44}
\definecolor{goldenrod18818934}{RGB}{188,189,34}
\definecolor{lightgray204}{RGB}{204,204,204}
\definecolor{mediumpurple148103189}{RGB}{148,103,189}
\definecolor{steelblue31119180}{RGB}{31,119,180}

\begin{axis}[
  name=TP1,
legend cell align={left},
legend style={
  fill opacity=0.8,
  draw opacity=1,
  text opacity=1,
  at={(0.97,0.03)},
  anchor=south east,
  draw=lightgray204,
  font=\scriptsize,
  /tikz/row sep=-5pt,
  inner ysep=-1pt
},
scale only axis,
width=0.24\textwidth,
height=0.24\textwidth,
log basis x={10},
log basis y={10},
tick align=outside,
tick pos=left,
title={(a)},
x grid style={darkgray176},
xlabel={$h$},
ylabel={Error},
xmin=0.0400529275138788, xmax=0.386351466975391,
xmode=log,
xtick style={color=black},
xtick={0.001,0.01,0.05,0.1,0.2,1,10},
xticklabels={
  $\mathdefault{10^{-3}}$,
  $\mathdefault{10^{-2}}$,
  $\mathdefault{5\times10^{-2}}$,
  $\mathdefault{10^{-1}}$,
  $\mathdefault{2\times10^{-1}}$,
  $\mathdefault{10^{0}}$,
  $\mathdefault{10^{1}}$
},
y grid style={darkgray176},
ymin=2.10482454942171e-05, ymax=1.23609864518465,
ymode=log,
ytick style={color=black},
ytick={1e-06,1e-05,0.0001,0.001,0.01,0.1,1,10,100},
yticklabels={
  $\mathdefault{10^{-6}}$,
  $\mathdefault{10^{-5}}$,
  $\mathdefault{10^{-4}}$,
  $\mathdefault{10^{-3}}$,
  $\mathdefault{10^{-2}}$,
  $\mathdefault{10^{-1}}$,
  $\mathdefault{10^{0}}$,
  $\mathdefault{10^{1}}$,
  $\mathdefault{10^{2}}$
}
]
\addplot [thick, \ColorPhiCluster, forget plot]
table {%
0.348529434 0.75039132
0.277453762 0.54498299
0.176627595 0.12844367
0.140629466 0.09943412
0.0886860942 0.04637355
0.0704808022 0.0220493899999998
0.0443994274 0.01443672
};
\addplot [thick, \ColorPhiFull, forget plot]
table {%
0.348529434 0.0167713300000001
0.277453762 0.000811289999999909
0.176627595 0.00092241000000004
0.140629466 0.000659469999999995
0.0886860942 0.000198200000000037
0.0704808022 0.000160119999999875
0.0443994274 5.41200000001574e-05
};
\addplot [thick, \ColorPhiRep, forget plot]
table {%
0.348529434 0.118623913333333
0.277453762 0.550288146666667
0.176627595 0.0287047533333333
0.140629466 0.119299263333333
0.0886860942 0.00459103333333344
0.0704808022 0.0270089383333332
0.0443994274 0.00103329666666663
};
\addplot [semithick, darkturquoise23190207, dash pattern=on 1.5pt off 2.475pt]
table {%
0.348529434 0.75039132
0.277453762 0.597363879190031
0.176627595 0.380282986832255
0.140629466 0.302778245761117
0.0886860942 0.190943056167768
0.0704808022 0.151746673417307
0.0443994274 0.0955929160747157
};
\addlegendentry{$h^1$}
\addplot [semithick, goldenrod18818934, dash pattern=on 1.5pt off 2.475pt]
table {%
0.348529434 0.75039132
0.277453762 0.475543352714903
0.176627595 0.192719646695888
0.140629466 0.12216914516839
0.0886860942 0.0485869835203949
0.0704808022 0.0306867260847565
0.0443994274 0.0121776536589837
};
\addlegendentry{$h^2$}
\addplot [semithick, mediumpurple148103189, dash pattern=on 1.5pt off 2.475pt]
table {%
0.348529434 0.0167713300000001
0.277453762 0.00846098596695586
0.176627595 0.00218285449764102
0.140629466 0.0011017374302447
0.0886860942 0.000276322155090128
0.0704808022 0.000138695265688487
0.0443994274 3.46721864252316e-05
};
\addlegendentry{$h^3$}
\end{axis}


%% file: TP_h_Theta10_runtime_v2.tex

\definecolor{darkgray176}{RGB}{176,176,176}
\definecolor{darkorange25512714}{RGB}{255,127,14}
\definecolor{forestgreen4416044}{RGB}{44,160,44}
\definecolor{lightgray204}{RGB}{204,204,204}
\definecolor{steelblue31119180}{RGB}{31,119,180}

\begin{axis}[
  name=TP2,
    at={(TP1.right of south east)},
    anchor=left of south west,
legend cell align={left},
legend style={
  fill opacity=0.8,
  draw opacity=1,
  text opacity=1,
  at={(0.03,0.97)},
  anchor=north west,
  draw=lightgray204,
  font=\footnotesize
},
scale only axis,
width=0.24\textwidth,
height=0.24\textwidth,
tick align=outside,
tick pos=left,
title={(b)},
x grid style={darkgray176},
xlabel={$r$},
ylabel={Time in s},
xmin=4.75, xmax=10.25,
xtick style={color=black},
y grid style={darkgray176},
ymin=-3.97008332335, ymax=85.95157479635,
ytick style={color=black}
]
\addplot [semithick, \ColorPhiFull]
table {%
5 0.117264747619629
6 0.352565288543701
7 1.21527886390686
8 4.25139284133911
9 15.9585065841675
10 62.4519958496094
};
\addlegendentry{$\Phi_{\mathrm{full}}$}
\addplot [semithick, \ColorPhiCluster]
table {%
5 0.522594690322876
6 0.955164551734924
7 4.94876384735107
8 8.98185253143311
9 36.6393089294434
10 81.8642272949219
};
\addlegendentry{$\Phi_{\mathrm{cluster}}$}
\addplot [semithick, \ColorPhiRep]
table {%
5 3.18179343
6 3.37285368
7 3.81678776
8 4.75617358
9 6.31195872
10 9.77859561
};
\addlegendentry{$\Phi_{\mathrm{Repulsor}}$}

\addplot[
  only marks,
  mark=*,
  mark size=2,
  \ColorPhiRep
]
coordinates {
  (10,28.1892317)
};

\addplot[
  only marks,
  mark=*,
  mark size=2,
  \ColorPhiFull
]
coordinates {
  (10,64.1002174)
};

\end{axis}

%% file: TP_h_Theta5_r10_speedup_v2.tex

\definecolor{darkgray176}{RGB}{176,176,176}
\definecolor{darkorange25512714}{RGB}{255,127,14}
\definecolor{forestgreen4416044}{RGB}{44,160,44}
\definecolor{lightgray204}{RGB}{204,204,204}
\definecolor{steelblue31119180}{RGB}{31,119,180}

\begin{axis}[
    at={(TP2.right of south east)},
    anchor=left of south west,
legend cell align={left},
legend style={
  fill opacity=0.8,
  draw opacity=1,
  text opacity=1,
  at={(0.03,0.97)},
  anchor=north west,
  draw=lightgray204,
  font=\footnotesize
},
scale only axis,
width=0.24\textwidth,
height=0.24\textwidth,
tick align=outside,
tick pos=left,
tick label style={font=\tiny},
title={\textbf{(c)}},
x grid style={darkgray176},
xlabel={\# of processors},
ylabel={Speedup},
xmin=-1.35, xmax=50.35,
xtick style={color=black},
y grid style={darkgray176},
ymin=-1.17165881005995, ymax=46.604835011259,
ytick style={color=black}
]
\addplot [thick, \ColorPhiFull]
table {%
1 1
4 3.70814108848572
8 7.41439771652222
12 11.2501821517944
24 22.2420921325684
48 44.433177947998
};
\addplot [thick, \ColorPhiCluster]
table {%
1 1
2 1.77999806404114
4 3.08023619651794
8 2.94229435920715
12 2.83749651908875
24 2.78297185897827
48 2.07086133956909
};
\addplot [thick, \ColorPhiRep]
table {%
1 1
2 1.873378892
4 3.582481615
8 6.027401796
16 9.801825875
24 11.87134051
48 10.35100480
};

\addplot[
  only marks,
  mark=*,
  mark size=2,
  \ColorPhiRep
]
coordinates {
  (48,5.921853982)
};

\addplot[
  only marks,
  mark=*,
  mark size=2,
  \ColorPhiFull
]
coordinates {
  (48,44.350602)
};

\end{axis}
\end{tikzpicture}

%% file: meshRedistribution.pdf_tex
\begingroup%
  \makeatletter%
  \providecommand\color[2][]{%
    \errmessage{(Inkscape) Color is used for the text in Inkscape, but the package 'color.sty' is not loaded}%
    \renewcommand\color[2][]{}%
  }%
  \providecommand\transparent[1]{%
    \errmessage{(Inkscape) Transparency is used (non-zero) for the text in Inkscape, but the package 'transparent.sty' is not loaded}%
    \renewcommand\transparent[1]{}%
  }%
  \providecommand\rotatebox[2]{#2}%
  \newcommand*\fsize{\dimexpr\f@size pt\relax}%
  \newcommand*\lineheight[1]{\fontsize{\fsize}{#1\fsize}\selectfont}%
  \ifx\svgwidth\undefined%
    \setlength{\unitlength}{595.27559055bp}%
    \ifx\svgscale\undefined%
      \relax%
    \else%
      \setlength{\unitlength}{\unitlength * \real{\svgscale}}%
    \fi%
  \else%
    \setlength{\unitlength}{\svgwidth}%
  \fi%
  \global\let\svgwidth\undefined%
  \global\let\svgscale\undefined%
  \makeatother%
  \begin{picture}(1,0.52380949)%
    \lineheight{1}%
    \setlength\tabcolsep{0pt}%
    \put(0,0){\includegraphics[width=\unitlength,page=1]{meshRedistribution.pdf}}%
  \end{picture}%
\endgroup%

%% file: oblateToSto.pdf_tex
\begingroup%
  \makeatletter%
  \providecommand\color[2][]{%
    \errmessage{(Inkscape) Color is used for the text in Inkscape, but the package 'color.sty' is not loaded}%
    \renewcommand\color[2][]{}%
  }%
  \providecommand\transparent[1]{%
    \errmessage{(Inkscape) Transparency is used (non-zero) for the text in Inkscape, but the package 'transparent.sty' is not loaded}%
    \renewcommand\transparent[1]{}%
  }%
  \providecommand\rotatebox[2]{#2}%
  \newcommand*\fsize{\dimexpr\f@size pt\relax}%
  \newcommand*\lineheight[1]{\fontsize{\fsize}{#1\fsize}\selectfont}%
  \ifx\svgwidth\undefined%
    \setlength{\unitlength}{591.71134889bp}%
    \ifx\svgscale\undefined%
      \relax%
    \else%
      \setlength{\unitlength}{\unitlength * \real{\svgscale}}%
    \fi%
  \else%
    \setlength{\unitlength}{\svgwidth}%
  \fi%
  \global\let\svgwidth\undefined%
  \global\let\svgscale\undefined%
  \makeatother%
  \begin{picture}(1,0.40650933)%
    \lineheight{1}%
    \setlength\tabcolsep{0pt}%
    \put(0,0){\includegraphics[width=\unitlength,page=1]{oblateToSto.pdf}}%
    \put(0.05027542,0.19493857){\makebox(0,0)[lt]{\lineheight{1.25}\smash{\begin{tabular}[t]{l}$t=0.1$\end{tabular}}}}%
    \put(0.25235895,0.19493857){\makebox(0,0)[lt]{\lineheight{1.25}\smash{\begin{tabular}[t]{l}$t=1.0$\end{tabular}}}}%
    \put(0.45351552,0.19493857){\makebox(0,0)[lt]{\lineheight{1.25}\smash{\begin{tabular}[t]{l}$t=2.0$\end{tabular}}}}%
    \put(0.64504908,0.19493857){\makebox(0,0)[lt]{\lineheight{1.25}\smash{\begin{tabular}[t]{l}$t=3.0$\end{tabular}}}}%
    \put(0.86186963,0.19493857){\makebox(0,0)[lt]{\lineheight{1.25}\smash{\begin{tabular}[t]{l}$t=4.0$\end{tabular}}}}%
    \put(0.05027542,0.00176438){\makebox(0,0)[lt]{\lineheight{1.25}\smash{\begin{tabular}[t]{l}$t=5.0$\end{tabular}}}}%
    \put(0.25235895,0.00176438){\makebox(0,0)[lt]{\lineheight{1.25}\smash{\begin{tabular}[t]{l}$t=6.0$\end{tabular}}}}%
    \put(0.45351552,0.00176438){\makebox(0,0)[lt]{\lineheight{1.25}\smash{\begin{tabular}[t]{l}$t=7.0$\end{tabular}}}}%
    \put(0.64504908,0.00176438){\makebox(0,0)[lt]{\lineheight{1.25}\smash{\begin{tabular}[t]{l}$t=8.0$\end{tabular}}}}%
    \put(0.86186963,0.00176438){\makebox(0,0)[lt]{\lineheight{1.25}\smash{\begin{tabular}[t]{l}$t=9.0$\end{tabular}}}}%
    \put(0,0){\includegraphics[width=\unitlength,page=2]{oblateToSto.pdf}}%
  \end{picture}%
\endgroup%

%% file: oblateToSto_glyphs_New.pdf_tex
\begingroup%
  \makeatletter%
  \providecommand\color[2][]{%
    \errmessage{(Inkscape) Color is used for the text in Inkscape, but the package 'color.sty' is not loaded}%
    \renewcommand\color[2][]{}%
  }%
  \providecommand\transparent[1]{%
    \errmessage{(Inkscape) Transparency is used (non-zero) for the text in Inkscape, but the package 'transparent.sty' is not loaded}%
    \renewcommand\transparent[1]{}%
  }%
  \providecommand\rotatebox[2]{#2}%
  \newcommand*\fsize{\dimexpr\f@size pt\relax}%
  \newcommand*\lineheight[1]{\fontsize{\fsize}{#1\fsize}\selectfont}%
  \ifx\svgwidth\undefined%
    \setlength{\unitlength}{395.99999135bp}%
    \ifx\svgscale\undefined%
      \relax%
    \else%
      \setlength{\unitlength}{\unitlength * \real{\svgscale}}%
    \fi%
  \else%
    \setlength{\unitlength}{\svgwidth}%
  \fi%
  \global\let\svgwidth\undefined%
  \global\let\svgscale\undefined%
  \makeatother%
  \begin{picture}(1,0.65286582)%
    \lineheight{1}%
    \setlength\tabcolsep{0pt}%
    \put(0,0){\includegraphics[width=\unitlength,page=1]{oblateToSto_glyphs_New.pdf}}%
    \put(0.15494312,0.3786879){\makebox(0,0)[lt]{\lineheight{1.25}\smash{\begin{tabular}[t]{l}$t=0.2$\end{tabular}}}}%
    \put(0.6784812,0.36770403){\makebox(0,0)[lt]{\lineheight{1.25}\smash{\begin{tabular}[t]{l}$t=2.0$\end{tabular}}}}%
    \put(0.15494312,0.00263637){\makebox(0,0)[lt]{\lineheight{1.25}\smash{\begin{tabular}[t]{l}$t=6.0$\end{tabular}}}}%
    \put(0.6784812,0.00376058){\makebox(0,0)[lt]{\lineheight{1.25}\smash{\begin{tabular}[t]{l}$t=6.0$\end{tabular}}}}%
    \put(0,0){\includegraphics[width=\unitlength,page=2]{oblateToSto_glyphs_New.pdf}}%
  \end{picture}%
\endgroup%

%% file: obstacle_NoSphere.pdf_tex
\begingroup%
  \makeatletter%
  \providecommand\color[2][]{%
    \errmessage{(Inkscape) Color is used for the text in Inkscape, but the package 'color.sty' is not loaded}%
    \renewcommand\color[2][]{}%
  }%
  \providecommand\transparent[1]{%
    \errmessage{(Inkscape) Transparency is used (non-zero) for the text in Inkscape, but the package 'transparent.sty' is not loaded}%
    \renewcommand\transparent[1]{}%
  }%
  \providecommand\rotatebox[2]{#2}%
  \newcommand*\fsize{\dimexpr\f@size pt\relax}%
  \newcommand*\lineheight[1]{\fontsize{\fsize}{#1\fsize}\selectfont}%
  \ifx\svgwidth\undefined%
    \setlength{\unitlength}{609.44401779bp}%
    \ifx\svgscale\undefined%
      \relax%
    \else%
      \setlength{\unitlength}{\unitlength * \real{\svgscale}}%
    \fi%
  \else%
    \setlength{\unitlength}{\svgwidth}%
  \fi%
  \global\let\svgwidth\undefined%
  \global\let\svgscale\undefined%
  \makeatother%
  \begin{picture}(1,0.52804492)%
    \lineheight{1}%
    \setlength\tabcolsep{0pt}%
    \put(0,0){\includegraphics[width=\unitlength,page=1]{obstacle_NoSphere.pdf}}%
    \put(0.57034013,0.26287215){\makebox(0,0)[lt]{\lineheight{1.25}\smash{\begin{tabular}[t]{l}$t=0.5$\end{tabular}}}}%
    \put(0.32013571,0.26287215){\makebox(0,0)[lt]{\lineheight{1.25}\smash{\begin{tabular}[t]{l}$t=0.3$\end{tabular}}}}%
    \put(0.82300574,0.00171308){\makebox(0,0)[lt]{\lineheight{1.25}\smash{\begin{tabular}[t]{l}$t=3.0$\end{tabular}}}}%
    \put(0.06993132,0.26287215){\makebox(0,0)[lt]{\lineheight{1.25}\smash{\begin{tabular}[t]{l}$t=0.1$\end{tabular}}}}%
    \put(0.82054451,0.26287215){\makebox(0,0)[lt]{\lineheight{1.25}\smash{\begin{tabular}[t]{l}$t=1.0$\end{tabular}}}}%
    \put(0.06879311,0.00171308){\makebox(0,0)[lt]{\lineheight{1.25}\smash{\begin{tabular}[t]{l}$t=1.5$\end{tabular}}}}%
    \put(0.32019732,0.00171308){\makebox(0,0)[lt]{\lineheight{1.25}\smash{\begin{tabular}[t]{l}$t=2.0$\end{tabular}}}}%
    \put(0.57160151,0.00171308){\makebox(0,0)[lt]{\lineheight{1.25}\smash{\begin{tabular}[t]{l}$t=2.5$\end{tabular}}}}%
    \put(0,0){\includegraphics[width=\unitlength,page=2]{obstacle_NoSphere.pdf}}%
  \end{picture}%
\endgroup%